\documentclass[journal,12pt,onecolumn,draftclsnofoot,]{IEEEtran}
\usepackage[utf8]{inputenc} 
\usepackage{amsmath,amsfonts}
\usepackage{algorithmic}
\usepackage{algorithm}
\usepackage{array}
\usepackage[caption=false,font=normalsize,labelfont=sf,textfont=sf]{subfig}
\usepackage{textcomp}
\usepackage{stfloats}
\usepackage{url}
\usepackage{verbatim}
\usepackage{graphicx}
\usepackage{cite}
\usepackage{color}
\usepackage{units} 
\usepackage{textgreek} 
\usepackage{enumitem} 
\usepackage[version=4]{mhchem} 
\newcommand\dc{\ensuremath{_\text{dc}}} 
\newcommand\pump{\ensuremath{_\text{p}}} 
\newcommand\signal{\ensuremath{_\text{s}}} 
\newcommand\idler{\ensuremath{_\text{i}}} 
\newcommand\micro{\ensuremath{\text{\textmu}}} 
\newcommand\crit[1][]{\ensuremath{_{\text{c}#1}}} 
\newcommand\eg{\textit{e.g.}} 
\newcommand\ie{\textit{i.e.}} 
\newcommand\rd{\ensuremath{\mathrm{d}}} 
\newcommand\id{\ensuremath{\,\mathrm{d}}} 

\begin{document}

\title{Modeling and Harmonic Balance Analysis of Superconducting Parametric Amplifiers for Qubit Read-out: A Tutorial\\}

\author{Daryoush Shiri,~\IEEEmembership{Senior Member,~IEEE}, Hampus Renberg Nilsson, Pavan Telluri, Anita Fadavi Roudsari,~\IEEEmembership{Member,~IEEE}, Vitaly Shumeiko, Christian Fager,~\IEEEmembership{Senior Member,~IEEE}, Per Delsing

\thanks{}}



\maketitle



\section{Introduction}

Parametric modulation of reactive elements in circuits for amplification dates back to WWI, when Alexanderson and others proposed magnetic amplifiers for radio transmitters \cite{Alexanderson1916,Mumford1960,Zenneck1920,vanderZiel1948}. The low noise figure of the first parametric amplifiers, as a result of using passive elements, was on par with the standard at the time\cite{Kotzebue1961}. Later, the quest for wide-band amplification while obviating the bandwidth-gain trade-off led to the proposal of traveling-wave parametric amplifiers (TWPA). The proposed designs were based on modulating the inductance or capacitance values of a nonlinear transmission line by a traveling, high amplitude pump, a process akin to nonlinear optical materials where the refractive index (or group velocity) is power-dependent \cite{Tien1958,Landauer1960,Cullen1958}. After high-quality and low-loss germanium and silicon varactors became available, TWPAs were developed in different labs. In 1958 Bell Laboratories implemented the first example of a TWPA \textcolor{black}{with 10\,dB gain and 200\,MHz bandwidth centered at 700\,MHz \cite{Kotzebue1961}}. Readers can find other examples of single-diode or traveling-wave type amplifiers based on varactors in the complete bibliography of Blackwell and Kotzebue's book \cite{Kotzebue1961}.
With the advent of semiconductor transistor amplifiers in the 1970s, the interest in varactor-based parametric amplifiers declined. However, interest in TWPAs based on superconductors was on the rise for low-noise amplification of faint signals in astronomy and quantum physics.
\textcolor{black}{The first experiments showing parametric amplification in superconductors were reported by Bura \cite{Bura1966} and Clorfeine \cite{Clorfeine1964} in the mid 1960s; they attributed the gain to the nonlinear inductance of their superconducting films modulated by the pump. Later, a parametric amplifier using a single Josephson junction (JJ) was reported by H. Zimmer in 1967 \cite{Zimmer1967}. In these experiments, the film or JJ was mounted on a face of a waveguide, and the amplified signal was reflected and diverted to the load by a circulator.} 
\textcolor{black}{The Manley-Row relations, gain, and gain-bandwidth product of a JJ-based amplifier were later derived in a theoretical paper by Peter Russer \cite{Russer1969}}. Thereafter, the physics community witnessed a speedy development of JJ-based low-noise parametric amplifiers \cite{Yu1975,Kanter75,Goodall1979,Aumentado2020}. Quantum noise-limited microwave amplifiers based on JJs are vital in, for example increasing the sensitivity of dark matter detectors \cite{Dark2019} and the fidelity of a quantum bit (qubit) read-out \cite{White2023}. 
Today, myriad designs based on JJs \cite{Roy2015,kaufman2023josephson, Miano2022}, superconducting quantum interference devices (SQUIDs) \cite{Castellanos-Beltran2008}, and superconducting nonlinear asymmetric inductane elements (SNAILs) \cite{Frattini2017} are reported for amplification of \textcolor{black}{microwave signals carrying the information about the state of a qubit.}

\begin{figure}[!t]
\centering
\includegraphics[width=6 in]{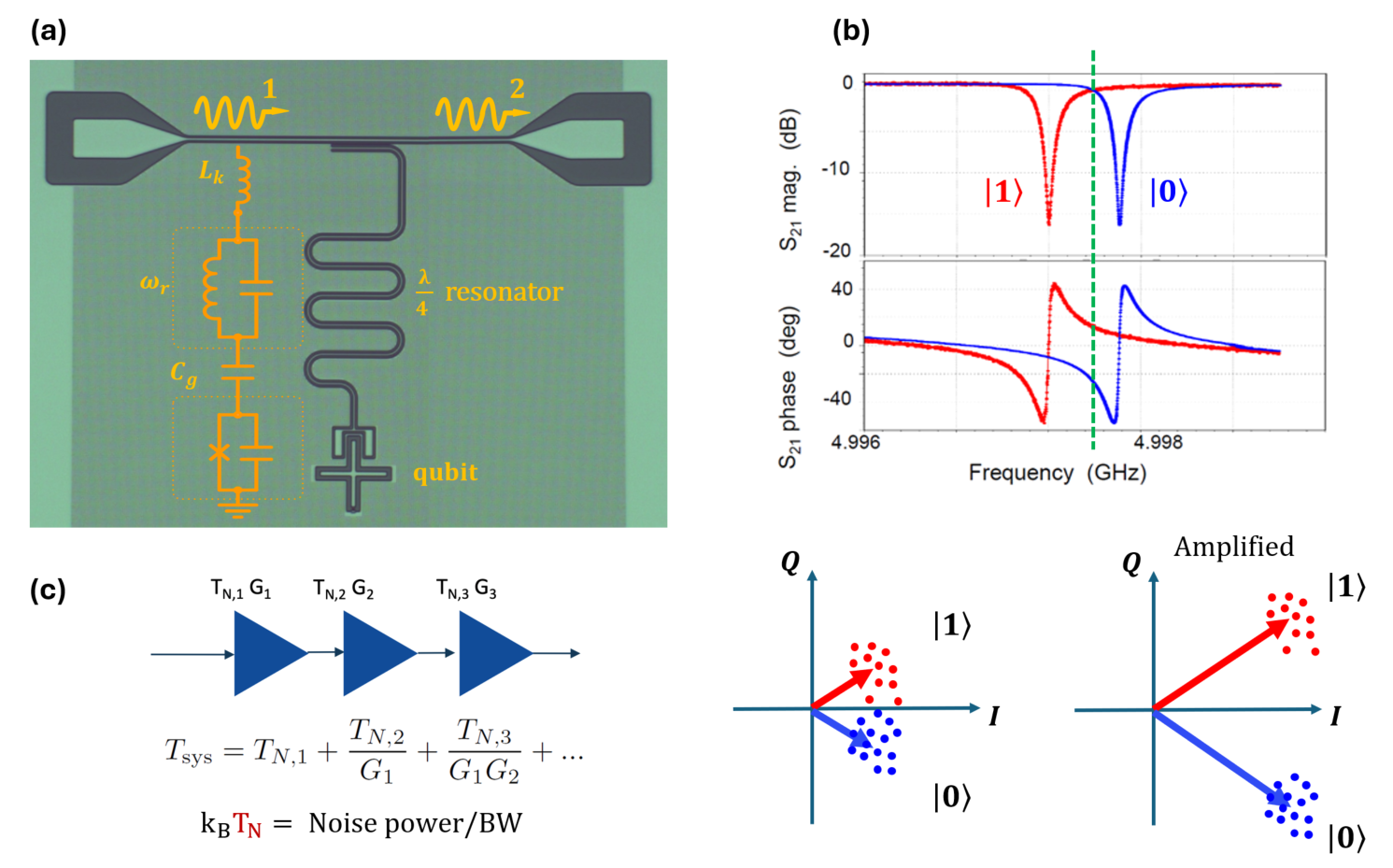}
\caption{\textcolor{black}{(a) A simplified diagram of a qubit coupled to a quarter-wavelength resonator to be read by a horizontal CPW feedline on top. The electric circuit model of the qubit, resonator, and coupling capacitors are shown on the leftmost part, where $C_{g}$ and $L_{k}$ are qubit-resonator coupling capacitance and resonator-feedline coupling inductance, respectively. $\omega_{r}$ is the resonance frequency of the read-out resonator. The transmission $S_{21}$ is measured by sending a continuous microwave signal at a constant frequency, as shown by the green dashed line in (b). The change of qubit state from $|0\rangle$ to $|1\rangle$ leads to a down-shift of the resonance frequency, as a result of which the phase of $S_{21}$ changes. The tip of the measured S-parameter vector on the IQ-plane is shown by dots (bottom left). The discrimination between states $|0\rangle$ and $|1\rangle$ and overcoming the noise is achieved by amplification (bottom right). (c) A chain of amplifiers to read the qubit state. $T_{N,i}$ and $G_i$ are the noise temperature of the amplifier $i$ and its power gain, respectively. The Friis formula shows why the first amplifier must be of a passive (parametric) type as the total noise temperature of the system is limited by $T_{N,1}$.}}
\label{fig_qubitS21}
\end{figure}

\textcolor{black}{Fig. \ref{fig_qubitS21}(a) shows a simplified portion of a silicon chip comprising a co-planar waveguide (CPW) feedline coupled to a quarter-wavelength resonator. The resonator is capacitively coupled to a qubit which is made of the parallel combination of a capacitor (cross-shaped) and a nonlinear inductor using a Josephson junction. This nonlinear resonator implements an artificial two-state atom or qubit with states 0 and 1. The physics of interaction of a qubit and a resonator resembles that of an atom inside an electromagnetic cavity \cite{Jaynes1963}. The resonance frequency of the $\lambda/4$-resonator (or cavity) is sensitive to the state of the atom or qubit inside the cavity as shown in Fig. \ref{fig_qubitS21}(b) \cite{Bardin2020}. Therefore, the qubit state can be measured by monitoring the transmission $S_{21}$ through the read-out CPW feedline. As an example, by sending a fixed frequency continuous microwave and measuring the in-phase (I) and quadrature (Q) components of the output signal, the state of the qubit can be mapped to the IQ-plane. However, to detect the qubit states with better fidelity, amplification is necessary. According to the Friis formula shown in Fig. \ref{fig_qubitS21}(c), the total noise temperature of the system is determined by the first amplifier in the amplification chain of the qubit read-out line \cite{Friis1944,Aumentado2020}. This mandates the use of a low-noise and high-gain amplifier at the first stage. As a result, amplification using active elements like transistors and Gunn diodes is not possible, and use of parametric amplifiers based on passive nonlinear reactive elements like inductors or capacitors is necessary.} The low noise figure of the parametric amplifiers results in high fidelity in discriminating the states 0 and 1 of the qubits \textcolor{black}{represented by vectors $|0\rangle=[1~~0]$ and $|1\rangle=[0~~1]$, respectively.} A gain of 15-20\,dB and bandwidth of 700\,MHz to multi-\,GHz with large tunability and low noise (less than a photon per unit bandwidth) have been reported \cite{Anita2023,White2023,Datta2022}. \textcolor{black}{ A total noise of around 1.3 photons (added noise of around 0.6 photons) has been measured and simulated in \cite{Salowmir2021,Yu1975}}.

Using the nonlinearity due to the kinetic inductance of a superconducting transmission line is another method for implementing TWPAs \cite{Kind2018,Kind2021,Sweetnam2022}. \textcolor{black}{The kinetic inductance originates from the inertia of superconducting electrons in response to the applied electric field. For a film of length $l$ and cross sectional area of $A$, $L_\mathrm{kin}=(l/A)\mu_{o}\lambda_{\mathrm{L}}^{2}$ where $\lambda_{\mathrm{L}}$ is the London penetration depth of a magnetic field into the film \cite{VanDuzer2008} and $\mu_{o}$ is the magnetic permeability of vacuum. Although kinetic inductance TWPAs are easy to fabricate, they have to be longer and have higher pump powers for the nonlinear interaction to take effect and obtain a higher gain \cite{OBrien2014}.}

Predicting the performance of parametric amplifiers and capturing the processes behind amplification are vital because they provide insight into ways to enhance the gain mechanism and avoid the processes that compete with the amplification such as \textcolor{black}{pump depletion and up-conversion as a result of inter-mixing of pump, signal, and idler}. This, in turn, reduces the time-to-fabrication cycle. However, the complex mechanisms of power conversion between pump harmonics or undesired pump depletion effects, and compression point of the amplifier cannot be captured by traditional small-signal methods like AC analysis and $S$-parameters. These methods are based on linearizing the circuit around the operating point by assuming a low amplitude stimulus applied around that point. The small-signal transfer functions like signal transmission ($S_{21}$) and reflection ($S_{11}$) are extracted for each given bias point and frequency \cite{Penfield1960,Kurokawa1965, Kundert2007,Kundert86}.  
However, in a strongly nonlinear circuit such as a TWPA, the injected high amplitude tone leads to the oscillation of the operating point. Furthermore, in the case of two or more input tones, the intermixing products make the mapping between the input and transmitted/reflected waves less straightforward and \textcolor{black}{render $S$-parameters meaningless}.
Extracting the spectral content of the amplifier output is possible by performing a time-domain transient analysis followed by a Fourier analysis. However, this requires a long simulation with a very small time-step. The former is because the initial reflections and ringing at the output of the circuit must settle down so that the circuit reaches a periodic steady state. The latter is because the maximum frequency of interest (\eg, the 5th harmonic of the pump) sets a limit on the time-step by virtue of the Nyquist criterion. Moreover, to extract the gain spectrum, the above process must be repeated for every given signal, pump frequency, and power level, which makes the simulation very time-consuming.    

\textcolor{black}{In addition to using circuit simulators, there have been attempts to analytically model TWPAs based on coupled-mode theory, input-output theory, and full quantum mechanical treatment of different photon-photon interactions in TWPAs. The coupled-mode theory starts with writing a nonlinear wave equation describing the propagation of the wave along the TWPA. Using a trial set of (ansatz) solutions as a sum of pump, signal, and idler modes, it is possible to extract the coupled differential equations describing the amplitudes of these components from which the gain, gain compression, and harmonic generation are extracted. \cite{OBrien2014}. TWPAs are promising candidates to create a non-classical state of light at the output (\eg, squeezed or minimum uncertainty photon fields, and entangled photons). These processes can be modeled using input-output theory in which the TWPA is treated as a continuous medium, and the input and output waves are second-quantized to obtain the Hamiltonian of the TWPA system \cite{Grimsmo2017}. Digression from the continuous medium assumption is done in \cite{vanderReep2019}. Here the Hamiltonian is found by second-qauntizing the magnetic and electric energies in the circuit that also includes the nonlinear potential energy term in Josephson junctions. The time evolution of the amplitude of different fields (now treated as quantum mechanical operators) is governed by the Schrödinger equation. From this, the same coupled-mode equations and gain formula which are described in \cite{OBrien2014, Grimsmo2017} are extracted. However, to make the equations tractable, only the first term in the Taylor expansion of the Josephson junction potential energy is taken into account. Also, it is assumed that the pump is a classical wave and that pump depletion does not occur. A full quantum mechanical approach (Hamiltonian) is presented in \cite{Haider2024, Yuan2023}, where all the processes -- like three-wave mixing, four-wave mixing, cross-phase modulation, self-phase modulation, dissipation, and added quantum noise -- of TWPAs are treated. The results agree with the experiments. The modeling of the nonlinear elements in \cite{Haider2024} (JJ, SQUID, SNAIL), called blackbox, is based on the polynomial expansion of the potential energy of these elements as a function of magnetic flux and keeping the most important terms in this expansion. However, as we will see in this article, by mathematically modeling (wiring) the nonlinear elements like JJ and SNAIL and using harmonic balance (HB), there is no need to make any assumptions about the strength of the nonlinearity or the nature of the pump.}

Harmonic balance (HB) is the most reliable and dominant method for large signal analysis of highly nonlinear microwave circuits. It is used to analyze circuits with periodic, quasi-periodic, and steady-state responses. The origin of this method goes back to 1930's when Galerkin, Krylov, and Bogoliubov proposed that the solution of a nonlinear system can be written as a sum of known functions \cite{Gilmore91-1}. The unknown coefficients in the sum are found by starting with a trial solution in the governing equations of the nonlinear system and solving an algebraic equation. If the known functions are sinusoidal, then they are called harmonics, and the method is called harmonic balance. \textcolor{black}{This is when the output of the system is} expected to have a steady-state periodic waveform (not necessarily sinusoidal). The modern versions of this method to solve nonlinear circuits (\eg, power amplifiers and varactor-based mixers) appeared in the 1960s-1980s \cite{Lipparini1981,Filicori1979,Nakhla76,Kundert86}. Nowadays it is also used by mechanical and civil engineers in structural stability analysis. \\
In the HB method, the circuit is split into a linear sub-circuit and a nonlinear sub-circuit. The elements in the linear section are expressed by their frequency domain admittance or impedance, and the analysis is done in the Fourier domain. The nonlinear section is solved in the time-domain by algebraic methods or by solving time-domain differential equations, depending on whether the circuit is quasi-static \cite{Maas2003}. Details of HB method and its fundamentals can be found in \cite{Filicori1979,Gilmore91-1,Gilmore91_2,Nakhla76,Maas2003} and a short summary for the case of a single harmonic input signal is presented in Appendix A. \\
The most important options to choose at the beginning of HB simulation are (a) the maximum number of harmonics, $k$, (b) the method of inverting the Jacobian matrix of error equation (Krylov vs. Direct methods \cite{Maas2003,Golub2013}), and (c) the time-step for the transient analysis step of HB. The Jocobain matrix of the error equation is shown in equation (\ref{eqn_8}) in Appendix A.\\   
Before HB simulation, the first step in the design of a parametric amplifier is the mathematical modeling of the nonlinear element. The nonlinear element in a superconducting parametric amplifier is a JJ or combinations of JJs as SNAILs or SQUIDs. All these elements generally work as nonlinear inductors whose value (the parameter) is a function of a bias current and/or an applied magnetic flux. An essential feature of JJ-based devices is that all higher order nonlinear effects are known from the Josephson relations \cite{VanDuzer2008}. \\
The modeling of JJs and SNAILs using the symbolic device definition (SDD) is discussed in section II, where the scattering parameter of a SNAIL is also calculated and compared with the analytic method. Section III presents the design process of a TWPA with JJs and its gain simulation and measurement. It also shows how in the HB method, the convergence is achieved using the information gained from the transient analysis of the amplifier. \textcolor{black}{In the same section, the experimental characterization of a JJ-TWPA architectures is compared with simulations.} In section IV, the gain spectrum of a SNAIL-TWPA is simulated with the HB method and compared with the experimental data. The agreement between the simulation and measurement proves the efficiency and reliability of the HB method. Section V is focused on the modeling of power exchange between the pump harmonics, a process which is detrimental to the gain of TWPAs. We show how the HB analysis results agree with those of the coupled-mode theory \cite{Hampus2023, Armstrong1962} without the necessity of solving any set of coupled differential equations. The insight obtained from this part leads to creative solutions to achieve TWPAs of higher gain and bandwidth.

\section{Symbolically Defined Model of Josephson Junction and SNAIL}
Equation-based modeling of nonlinear components is possible in modern circuit simulators like AWR Microwave Office, APLAC \cite{Kiviranta_IEEE}, and in general, using a Verilog A code. In this work, we use the symbolically defined device (SDD) in Keysight ADS, which allows direct mathematical modeling of a device in both large-signal and small-signal regimes. The mathematical functions relating the voltages and currents \textcolor{black}{of the device's ports} can be modeled implicitly using $f(I,V)=0$, or explicitly by giving the relation between the current and voltage of each element or port, $I=g(V)$, where $f(\cdot)$ and $g(\cdot)$ are arbitrary nonlinear functions that can be derived from the Josephson relations. The device can have as many ports as required. Here we show how JJs and SNAILs are modeled as a black box using SDD. \textcolor{black}{By blackbox we mean that the equations describing the operation of each device are mathematically implemented by wiring the circuit components in the simulator. This is akin to modeling a BJT transistor using diodes, controlled sources, resistors, and capacitors to implement the Gummel-Poon equations. Furthermore, we do not need to do any polynomial expansion or make any assumptions about the strength of the nonlinearity in JJs or SNAILs \cite{Haider2024, Hampus2023}.}      
\subsection{Josephson Junction (JJ)}
\begin{figure}[!t]
\centering
\includegraphics[width=3.5in]{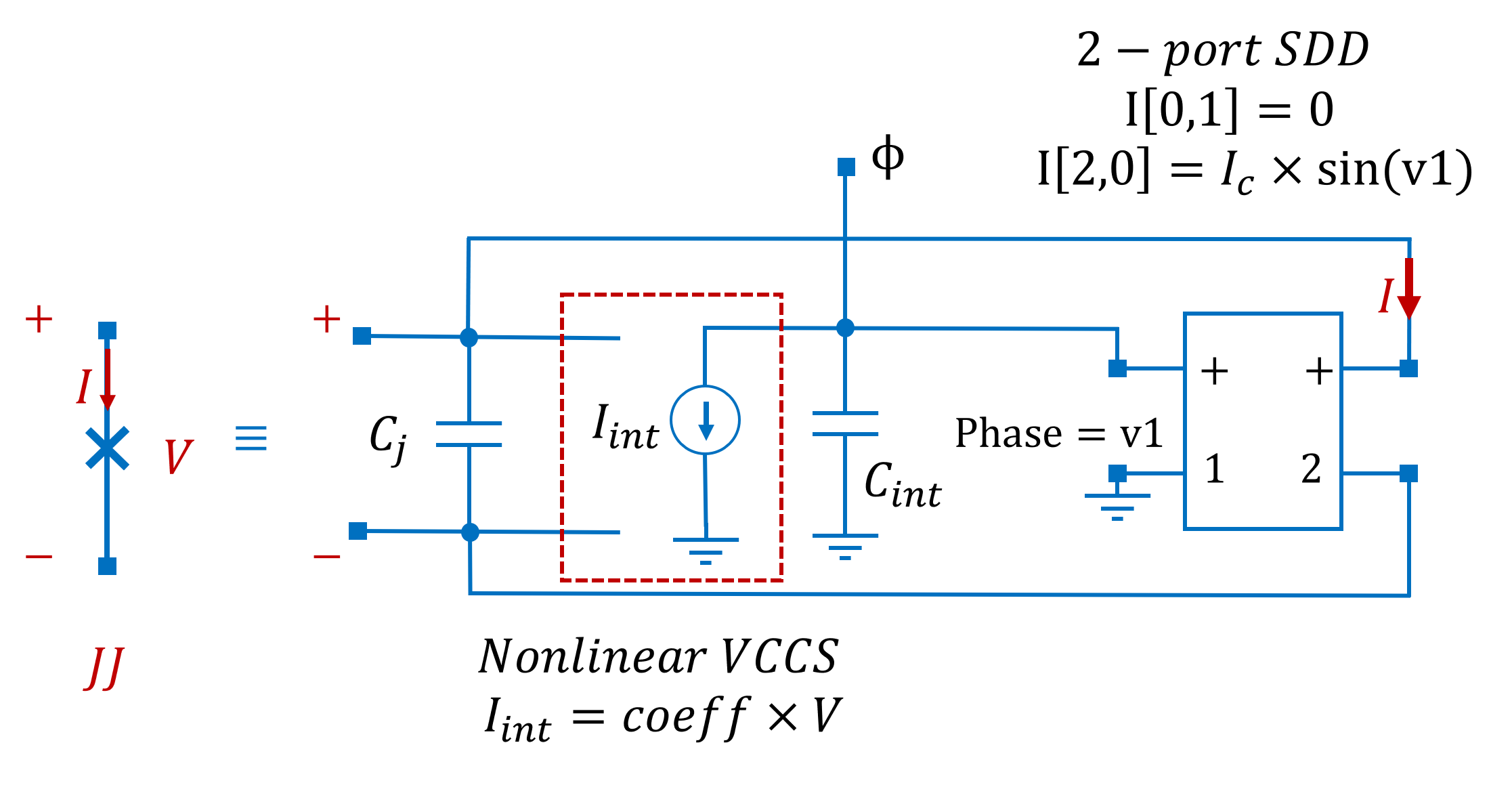}
\caption{The circuit symbol (left) and the schematics of the mathematical model (right) of a JJ's behavior based on equation (\ref{eqn_9}).}
\label{fig_2JJ}
\end{figure}

The Josephson junction is composed of a sandwich of metal-insulator-metal layers. At temperatures below the critical temperature ($T<T\crit$), the metal becomes superconducting, where the electrons pair up into so-called Cooper pairs. For aluminum, the critical temperature is typically around $T\crit= 1.2 K$. The wave function of the Cooper pair condensate is a complex quantity whose norm squared is equal to the density of the Cooper pairs. The phase difference between the wave functions of the Cooper pairs on the two JJ terminals is denoted by $\phi$. The voltage is the time derivative of this phase, which is proportional to the magnetic flux, $\Phi=({\Phi_0}/{2\pi})\phi$, where $\Phi_0 = \unit[2.07 \times 10^{-15}]{Wb}$ is the magnetic flux quantum. \textcolor{black}{The current and voltage of a Josephson junction in the superconducting state are related through Josephson relations, as shown below \cite{VanDuzer2008}.}
\begin{equation}
\label{eqn_9}
\begin{array}{cc}
I = I\crit \sin(\phi) \quad \text{for} \quad I < I\crit \\
V = \frac{\rd\Phi}{\rd t}=\frac{\Phi_0}{2\pi}\frac{\rd\phi}{\rd t} \Rightarrow \phi=\frac{2\pi}{\Phi_0}\int V\id t. 
\end{array} 
\end{equation}
\noindent $I\crit$ is the critical current. If the current is higher than this value, the JJ exits the superconducting state and becomes dissipative, and its behavior approaches that of a normal resistor with resistance $R_\mathrm{N}$. Note that this resistor is fixed only if the voltage applied to the JJ is larger than $2\Delta(T)/e$ and $T=0$, where $\Delta(T)$ is the energy to break the Cooper pairs and $e$ is the electronic charge \cite{VanDuzer2008}. In that case, $R_\mathrm{N}=\pi\Delta(T=0)/(2eI\crit)$. Otherwise, a nonlinear voltage- and temperature-dependent resistor, $R_\mathrm{N}(V, T)$, must be added in parallel to the model shown in Fig. \ref{fig_2JJ}. As we always work in the ($I<I\crit$) and ($T=0$) regime, modeling of the JJ's dissipative operation is not necessary. 
The potential energy stored in the JJ, $U$, is found from
\begin{equation}
\label{eqn_EJ4JJ}
U = \int V\cdot I \mathrm{d}t = -E_\mathrm{J}\cos(\phi), 
\end{equation}
where $E_\mathrm{J}={\Phi_{0}I\crit}/{2\pi}$ is called the Josephson energy. The Josephson relations mentioned in equation (\ref{eqn_9}) show that a JJ behaves like a nonlinear inductor, and its value is controlled by the bias current $I$. The nonlinear inductance is found from ${1}/{L_\mathrm{J}} = {\partial^2 U}/{\partial\Phi^2}$. \textcolor{black}{For the simple potential given in equation (\ref{eqn_EJ4JJ}), this differentiation is equivalent to writing $V=L_\mathrm{J}\mathrm{d}I/\mathrm{d}t$, which yields:}
\begin{equation}
\label{eqn_10}
L_\mathrm{J} = \frac{L_{\mathrm{J}0}}{\sqrt{1-(\frac{I}{I\crit})^2}}, 
\end{equation}
\textcolor{black}{where $L_{\mathrm{J}0}={\Phi_0}/(2\pi I\crit)$ is the zero-bias inductance.} To implement the JJ model, it is necessary to emulate the quantum mechanical phase mathematically. From the Josephson voltage-phase relation in equation (\ref{eqn_9}), the phase is found by integrating the input voltage. An ideal integrator is implemented by injecting a current into a capacitor. Therefore a nonlinear voltage-to-current converter (NonlinVCCS) is used to convert the voltage to current, and then it is injected into a capacitor (Fig. \ref{fig_2JJ}). The coefficient and the capacitor values are chosen to obtain the phase quantity with the correct dimensions in radians. With $C_\mathrm{int} = \unit[1]{pF}$ (any value is possible), the coefficient in NonlinVCCS is $ C_\mathrm{int}\times({2\pi}/{\Phi_0})=3038.5349$. The phase is then fed to the input port of a \textit{two-port} SDD. To guarantee an infinite input impedance at the input port (to avoid loading), we set the current of the input port to zero by setting $I[1,0]=0$. The output port current is then set to $I[2,0]=I\crit \sin(\phi)$. The output terminals of the two-port SDD are connected to both electrodes of the JJ to guarantee that this is the current that passes through the JJ. The physical capacitance of the metal-insulator-metal junction, $C_\mathrm{j}$, is added in parallel to the input ports, as shown in Fig. \ref{fig_2JJ}. The typical value for a JJ capacitance made of aluminum (\ie, a sandwich of Al-\ce{AlO_x}-Al) is about $C_\mathrm{j} = \unit[6-8]{fF}$. \textcolor{black}{This is an approximation based on post-fabrication measurement of the junction area, dielectric constant of aluminum oxide, and its thickness observed using transmission electron microscopy (TEM), which is about two nanometers.} Note that the ground terminals in the model are only mathematical zero reference voltages; they do not correspond to any physical ground terminal in the real device. 

\begin{figure}[!t]
\centering
\includegraphics[width=3.5in]{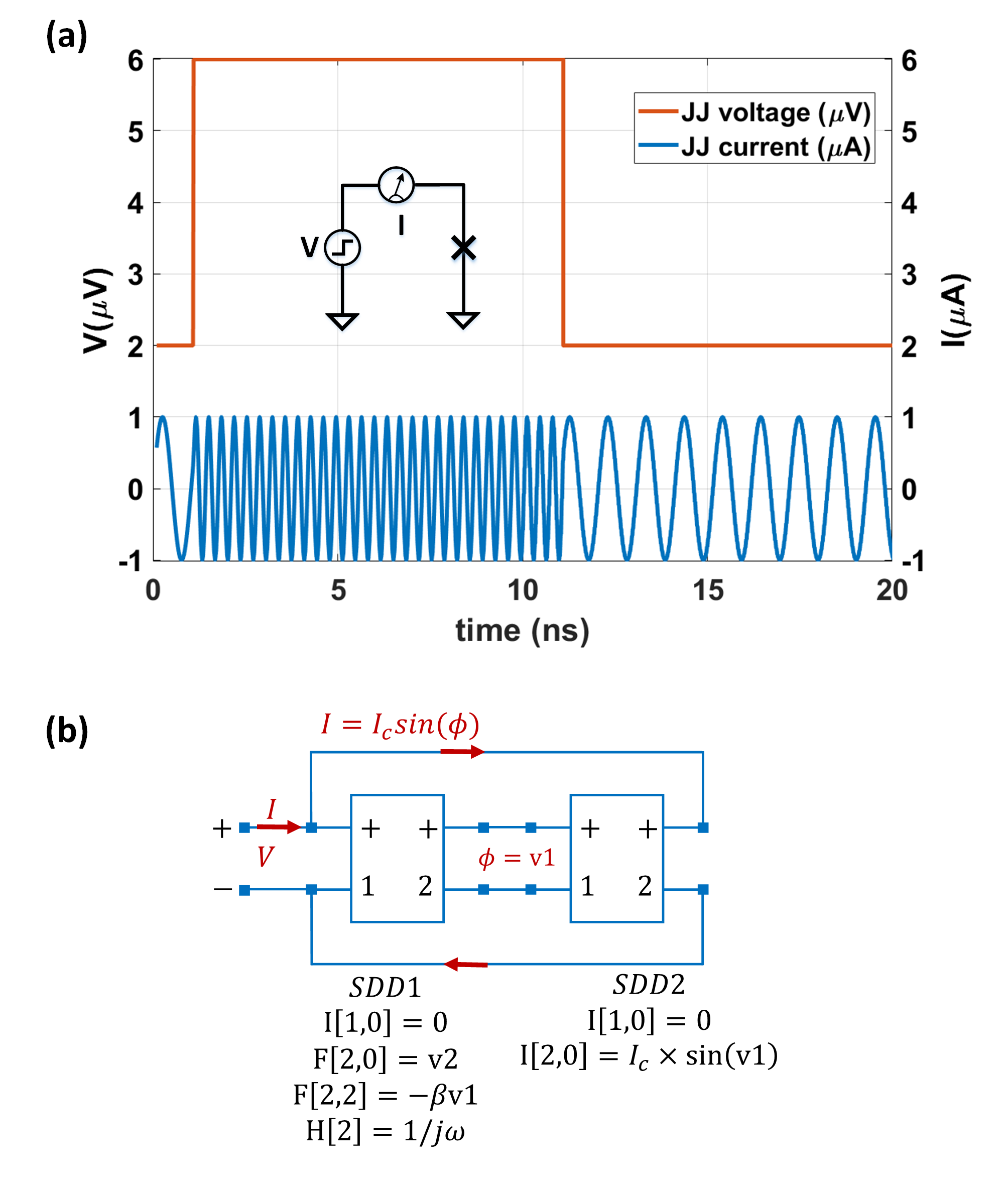}
\caption{(a) Simulated JJ current in response to an applied voltage pulse. (b) The JJ model based on two SDDs, where the first SDD implements the integrator in the frequency domain to create the phase, $\phi$.}
\label{fig_2SDD}
\end{figure}

The first simulation to check the correct operation of the JJ model is to apply a constant dc voltage, $V\dc$, in the time-domain and then observe the current oscillations. In this topology, the JJ works as a frequency modulator or a voltage-controlled oscillator circuit if a piecewise constant dc voltage is applied, as shown in Fig. \ref{fig_2SDD}(a). This is because the frequency of the current is proportional to the applied dc voltage. The frequency of the current oscillations [using equation (\ref{eqn_9})] is $f={V\dc}/{\Phi_0}$. In Fig. \ref{fig_2SDD}(a), the current frequency is $f = \unit[0.966]{GHz}$ and $f = \unit[2.898]{GHz}$, when the input voltage is $V\dc= \unit[2]{\mu V}$  and $V\dc=\unit[6]{\mu V}$, respectively.
Variation of the dc bias current leads to a change of inductance according to equation (\ref{eqn_10}). This can be tested by AC or $S$-parameter analysis of a single JJ and a parallel capacitor as an LC low-pass filter. By observing the change of bandwidth ($\omega_{\unit[-3]{dB}}$) as a function of the bias current, the inductance change in response to the current according to equation (\ref{eqn_10}) can be confirmed.

Note that the first integrator in Fig. \ref{fig_2JJ} can also be modeled using an SDD in implicit mode as shown in Fig. \ref{fig_2SDD}(b). Writing $F[2,0]=v_2$ and $F[2,2]=-\beta v_1$ enforces this equality: $v_2(\omega) =\beta H(\omega)v_2(\omega)$, where $\beta={2\pi}/{\Phi_0}$. The second index, 2 in  $F[2,2]$, tells the SDD to use a customized weighting function, $H(\omega)$, which is given by $H[2]=1/(j\omega)$. This means that in the frequency domain, the output voltage (phase) is the integral of the input voltage.

\subsection{Superconducting Nonlinear Asymmetric Inductance eLement (SNAIL)}

\begin{figure*}[!t]
\centering
\subfloat[]{\includegraphics[width=2.5in]{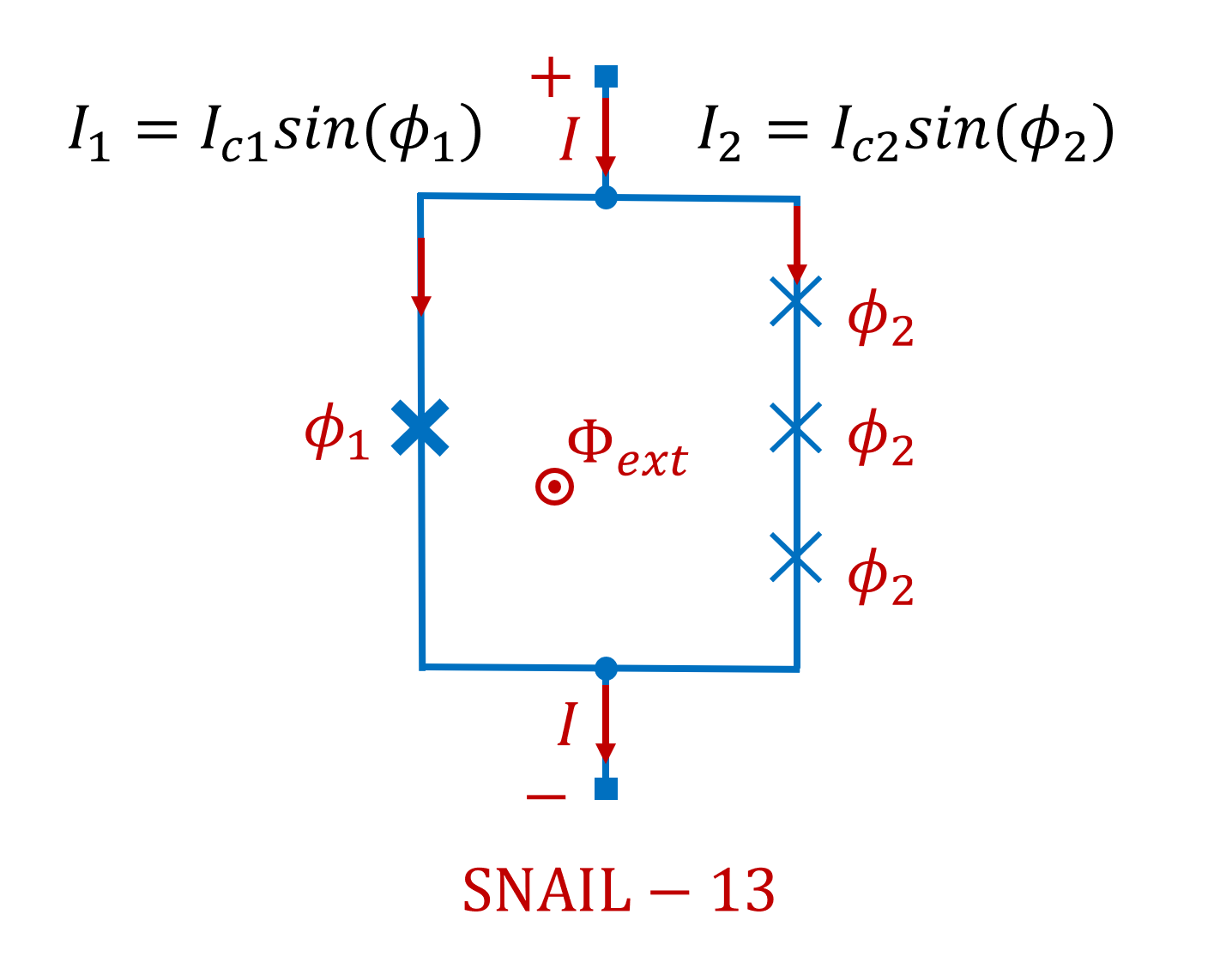}%
\label{fig_first_case}}
\hfil
\subfloat[]{\includegraphics[width=4in]{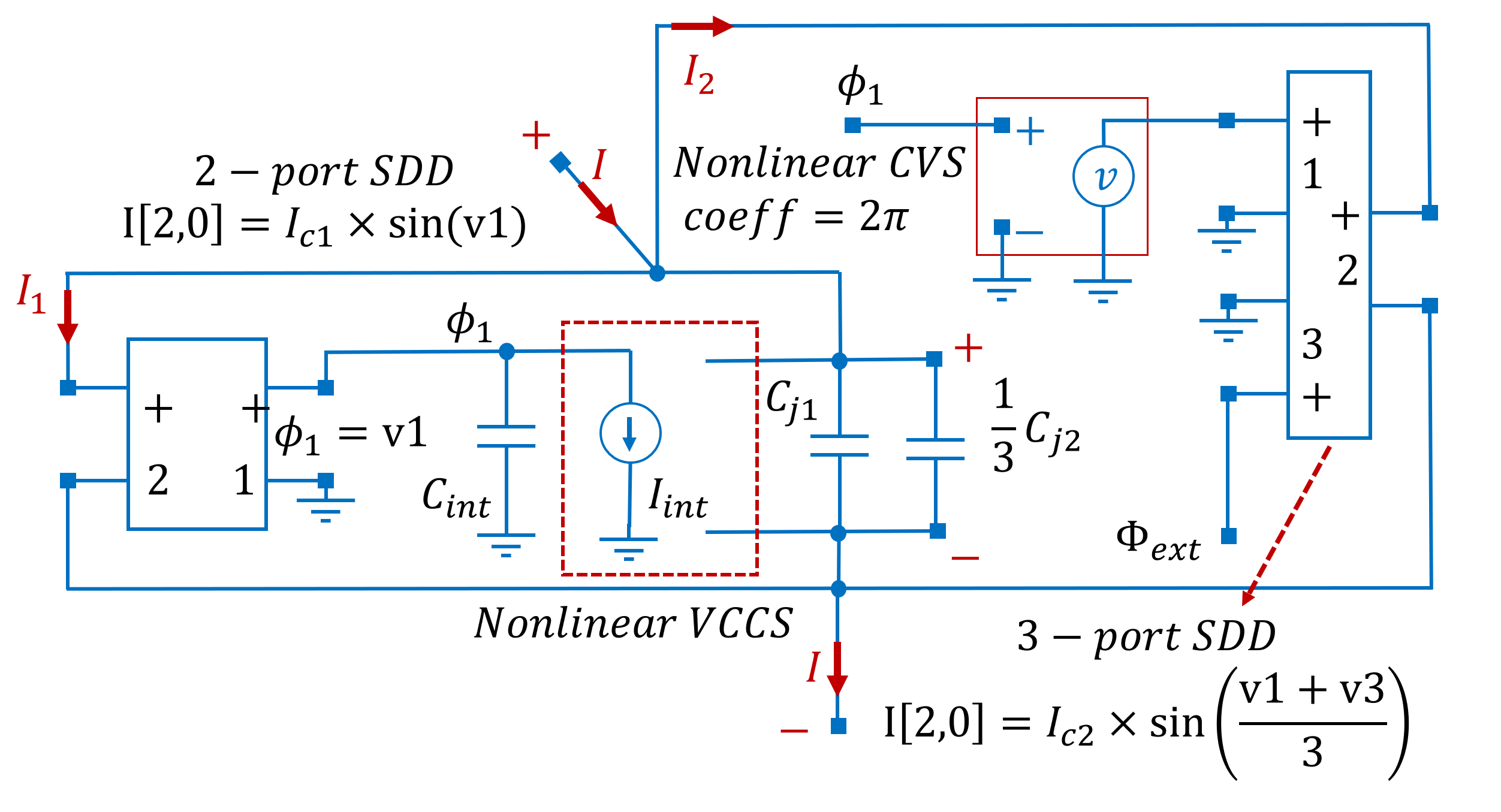}%
\label{fig_second_case}}
\caption{(a) Circuit topology of a SNAIL with three JJs in branch two or $N=3$. (b) Implementing the SNAIL model based on equations (\ref{eqn_9}) and (\ref{eqn_14}) by using the SDD and nonlinear controlled sources. Note that in general, $N$ can be different from 3 (\eg, for a SQUID, use $N=1$ and $\alpha=1$).}
\label{fig_3SNAIL}
\end{figure*}

\textcolor{black}{While it seems natural to discuss the model of a symmetric SQUID (\ie, a loop made of one Josephson junction in each branch), in this section, we actually begin by constructing a generalized SNAIL model. Then we demonstrate that, in a special case, it can be converted into a SQUID.} The tunability of the device inductance using an applied dc magnetic flux is attractive for amplifiers based on a 3-wave mixing mechanism \cite{Frattini2017,Hampus2023,Anita2023, Haider2024} as will be discussed in section IV. The circuit topology of a SNAIL element with three JJs in one branch is shown in Fig. \ref{fig_3SNAIL}(a). The right branch (branch 2) in Fig. \ref{fig_3SNAIL}(a) is composed of $N=3$ identical JJs. The Josephson energy of these three JJs is also $\alpha$ times that of the single junction in the left branch (branch 1). This asymmetry entails the following relation between critical currents of the two branches, which is $I\crit[1]=\alpha I\crit[2]$. Starting from KCL, we can write:
\textcolor{black}{
\begin{equation}
\label{eqn_11}
I = I_1 + I_2, \qquad\text{with}\qquad I_1 = I\crit[1] \sin(\phi_1) \qquad\text{and}\qquad I_2 = I\crit[2] \sin(\phi_2),
\end{equation}}
where $I$ is the total current passing through the SNAIL. Since the junctions on branch 2 are of the same size, the phase change across each one is the same. The total phase change summed around the loop, including the one due to the winding external flux, $\Phi_\text{ext}$, must be equal to zero, therefore:

\begin{equation}
\label{eqn_12}
\phi_1 - N\phi_2 + \frac{2\pi \Phi_\text{ext}}{\Phi_0}= 0, 
\end{equation}
from which (after setting ${\pi \Phi_\text{ext}}/{\Phi_0}= F$) we have, 

\begin{equation}
\label{eqn_13}
\phi_2 = \frac{2F+\phi_1}{N}.
\end{equation}

\begin{figure}[!t]
\centering
\includegraphics[width=3.5in]{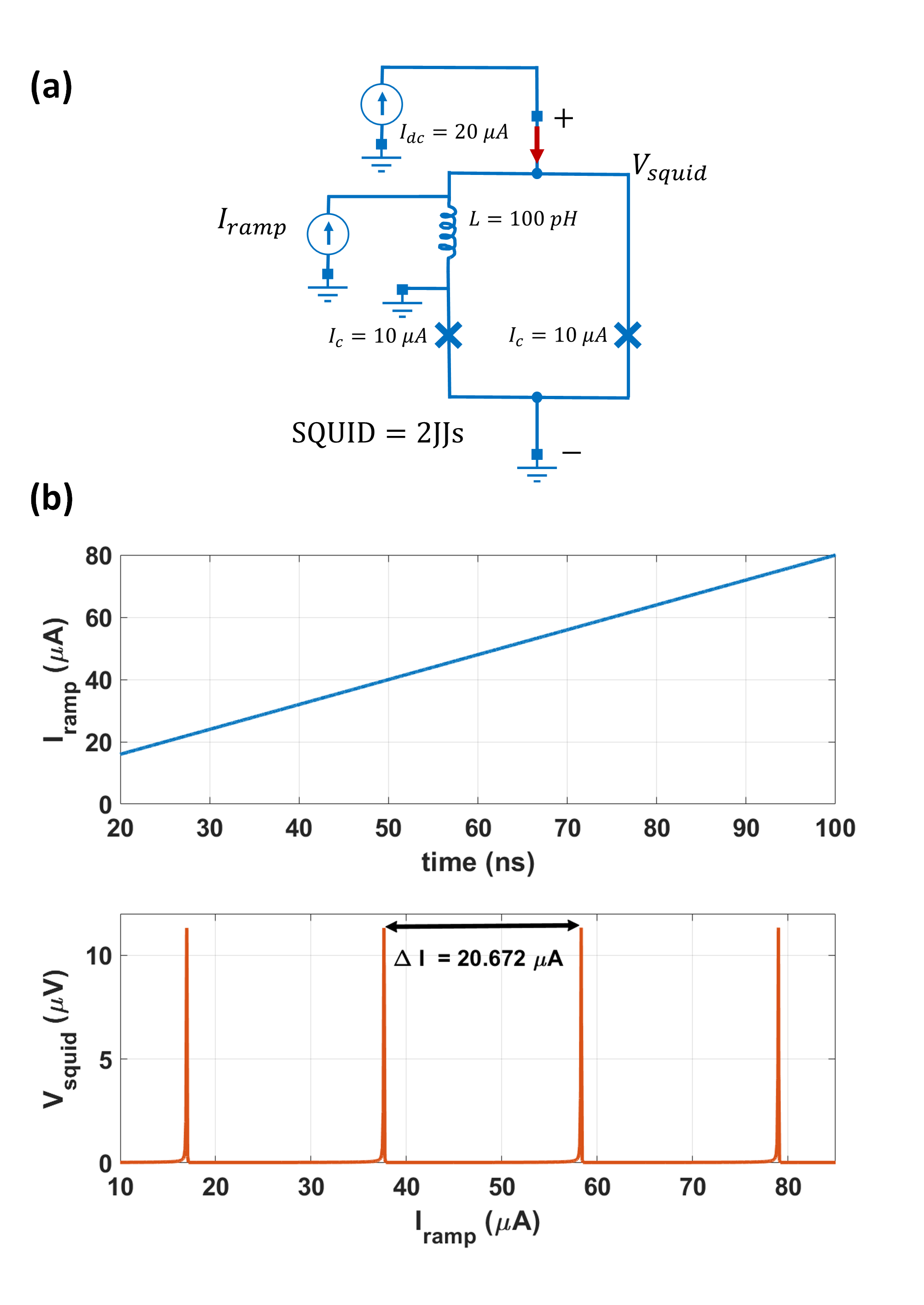}
\caption{\textcolor{black}{(a) Simulation setup for a SQUID to show the flux quantization. (b) The injected current ramp (top panel) and the SQUID voltage versus the current value (bottom panel). The pulses are almost $20.7 \mu A$ apart, showing a quantum of flux has wound the loop (\ie, $\Phi_0=\Delta I L$).}}
\label{SQUID}
\end{figure}

Note that in Fig. \ref{fig_3SNAIL} we have $N=3$. Equation (\ref{eqn_12}) mathematically guarantees the quantization of flux. \textcolor{black}{But the quantization of flux arises naturally as well if the SNAIL in Fig. \ref{fig_3SNAIL} is built by individual JJ black boxes modeled as in Fig. \ref{SQUID}. Before continuing, let's stop here and see how flux quantization occurs naturally if a SNAIL is modeled by two parallel JJ models. This specialized SNAIL with one JJ on each branch is called a SQUID. 
Fig. \ref{SQUID}(a) shows the simulation setup. It is assumed that the JJs are equal and each one has $ I\crit = 10\,\mu A$. The total bias current entering the loop is $I_\mathrm{dc}= 20\,\mu A$. A ramp current is injected into the left branch of the SQUID loop which is assumed to have $100~pH$ inductance. The voltage on the SQUID is zero except when the winding flux reaches one quantum of flux, $\Phi_0$. In this case, as shown in Fig. \ref{SQUID}(b), a pulse is generated so that the area under the pulse is $\Phi_0$. Plotting the SQUID voltage versus the ramp current shows that the periods of the pulses are equal to one flux quantum, because $\Delta I\times L = \unit[20.672]{\mu A} \times \unit[100]{pH} \approx 2.07 \times \unit[10^{-15}]{W} = \Phi_0$.}

Instead of juxtaposing individual JJ models, the SNAIL model (Fig. \ref{fig_3SNAIL}) is implemented by building two nonlinear current sources corresponding to each branch in equation (\ref{eqn_11}) and joining them together to build the total current $I$. Fig. \ref{fig_3SNAIL}(b) shows the model implemented in ADS. The left branch uses a two-port SDD to implement $I_1 = I\crit[1]\sin(\phi_1)$. The phase $\phi_1$ is created by integrating and scaling the voltage drop $V$ of the JJ (the leftmost branch of the SNAIL), as mentioned before for the single JJ. 
The current of the rightmost branch ($N$ JJ’s), the mathematical variable of phase $\phi_1$, and $2F$ are created and added together as two high-impedance voltage ports, $v_1$ and $v_3$, in the \textit{three-port} SDD. The rightmost port (port 2 or output) is then a sinusoidal function of the inputs, scaled by $I\crit[2]$, \ie,

\begin{equation}
\label{eqn_14}
I_2 = I\crit[2]\sin\left(\frac{2F+\phi_1}{N}\right).
\end{equation}

To simulate a symmetric SQUID which is a SNAIL with one JJ in each branch, it suffices to make $N=1$ and assume $I\crit[1]=I\crit[2]$ (\ie, $\alpha=1$). The physical capacitance of junctions can be added in parallel to their corresponding JJ voltage ports as shown by $C_{\mathrm{j}1}$ and $\frac{1}{3}C_{\mathrm{j}2}$. Note that the three series capacitors in Fig. \ref{fig_3SNAIL}(b), which model the physical capacitance of three JJs on the right branch, are grouped in one capacitor. Since we always assume that the circuit operates at a superconducting state, the normal resistances of the JJs, $R_\mathrm{N}(V,T)$, are not added to the SNAIL model.   

The equivalent inductance of the SNAIL is found from ${1}/{L_\text{tot}} = {\partial^2 E_\text{tot}}/{\partial\Phi^2}$, where $E_\text{tot}$ is the total Josephson (potential) energy of the SNAIL \textcolor{black}{[similar to equation (\ref{eqn_EJ4JJ})]} and it is written as:

\begin{equation}
\label{eqn_Etot}
E_\text{tot} = -E_{\mathrm{J}1}\cos(\phi_1) - NE_{\mathrm{J}2}\cos \left( \frac{2F+\phi_1}{N} \right )
\end{equation}
$E_{\mathrm{J}1}$ and $E_{\mathrm{J}2}$ are the Josephson energies of the JJs on the left and right branches of the SNAIL, respectively. From the above, it can be shown that $L_\text{tot}$ is
\begin{equation}
\label{eqn_15}
\frac{1}{L_\text{tot}} = \frac{1}{L_{\mathrm{J}0,\text{left}}}\cos(\phi_1) + \frac{\alpha}{NL_{\mathrm{J}0,\text{left}}}\cos\left(\frac{2F+\phi_1}{N}\right). 
\end{equation}

The individual inductance of the left branch at zero bias, $L_{\mathrm{J}0,\text{left}}$, has a similar form to that of equation (\ref{eqn_10}). Equation (\ref{eqn_15}) shows that the inductance is a function of the externally applied magnetic flux $F$. For analytic calculations, the value of $\phi_1$ is found by solving equation (\ref{eqn_11}) for a given input current $I$. \\
Note that the model of a SNAIL based on equations (\ref{eqn_12}) and (\ref{eqn_14}) generates the nonlinear inductance inherently and thus solving the extra equations (\ref{eqn_11}) and (\ref{eqn_15}) is not necessary.
The phase $\phi_1$ is also accessible in the SDD-based models as an extra node, \textcolor{black}{see Fig. \ref{fig_3SNAIL}(b)}. This is helpful in the design/simulation of logic gates based on quantum flux, for example, rapid single-flux quantum (RSFQ) \cite{VanDuzer2008} gates, when counting the number of flux quanta in the SNAIL or SQUID loops is necessary. The value of the phase node \textcolor{black}{shows how many flux quanta are carried by each voltage pulse, as the flux is the area under the voltage pulse,} see the Josephson relations and Fig. \ref{SQUID}(b). 

\begin{figure}[!t]
\centering
\includegraphics[width=17cm]{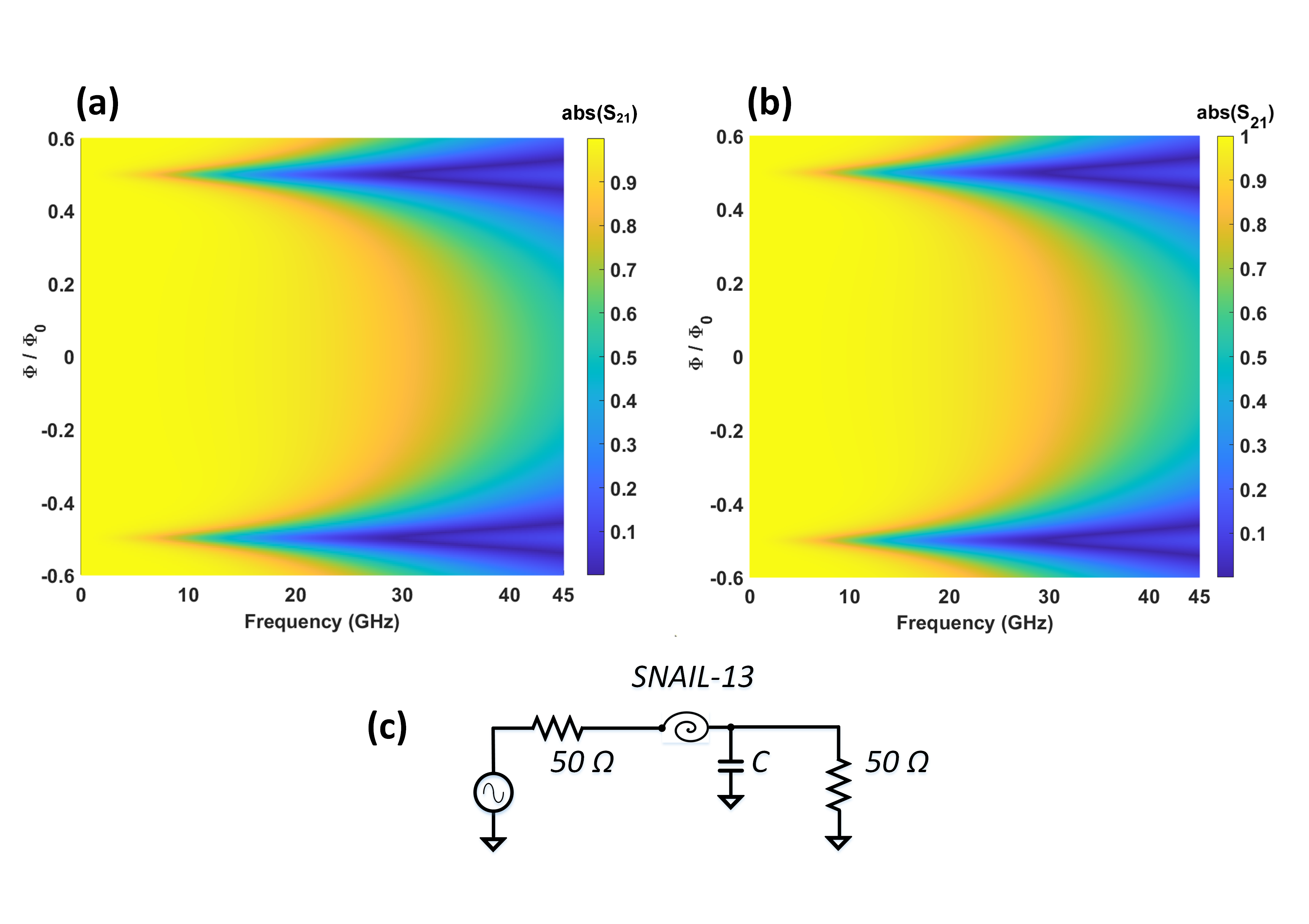}
\caption{(a) The magnitude of the transmission $S_{21}$, versus an external magnetic flux, based on analytic extraction from the ABCD matrix. (b) The same data from $S$-parameter analysis in ADS using the SDD-based model of a SNAIL-13. The flux is normalized to the quantum of flux, $\Phi_0$. $S_{21}$ changes periodically with the applied flux. \textcolor{black}{(c) The schematic of the LC low-pass filter, which later on is called the unit cell of the amplifier in section IV.}}
\label{fig_S21snail13}
\end{figure}

We conclude this section by showing the $S$-parameter analysis of an LC low-pass filter (LPF) where the inductor is made of a SNAIL-13. The index 13 means one and three JJ's in the left and right branches of the SNAIL, respectively (\ie, $N=3$). The right electrode of the SNAIL is connected to the $C=\unit[100]{fF}$ capacitor (that is shunted to ground) [see \textcolor{black}{Fig. \ref{fig_S21snail13}(c)]}. The $S_{21}$ of the LPF is found using both the numerical ABCD matrix method \cite{Collin_MW} and $S$-parameter analysis using the SDD model of the SNAIL. The results of both methods are identical. \textcolor{black}{However, the ABCD method involves solving two nonlinear equations to find the current and the quantum mechanical phase [equations (\ref{eqn_11}) and (\ref{eqn_12})] in an iterative manner for each given value of the external magnetic flux $F$. Thereafter, the value of the SNAIL inductance is found from equation (\ref{eqn_15}) and a T-network ABCD matrix is formed for SNAIL, C, and a short before the $\unit[50]{\Omega}$ load impedance in Fig. \ref{fig_S21snail13}(c). The S-parameters are extracted from the ABCD matrix \cite{Collin_MW}. In the general case of terminating the unit cells to non-$\unit[50]{\Omega}$ impedance, the conversion formula in \cite{Frickey1994} is used. Note that in ADS, the SNAIL is mathematically modeled (wired) as a circuit element and S-parameter simulation does not involve solving any nonlinear equations for a given external flux.} 
The transmission coefficient, $S_{21}$, varies with the applied magnetic flux with a periodicity of one flux quantum $\Phi_0$. This is evident from the current-phase dependence in equation (\ref{eqn_15}). 

\section{JJ-based TWPA}
\begin{figure}[!t]
\centering
\includegraphics[width=3.25in]{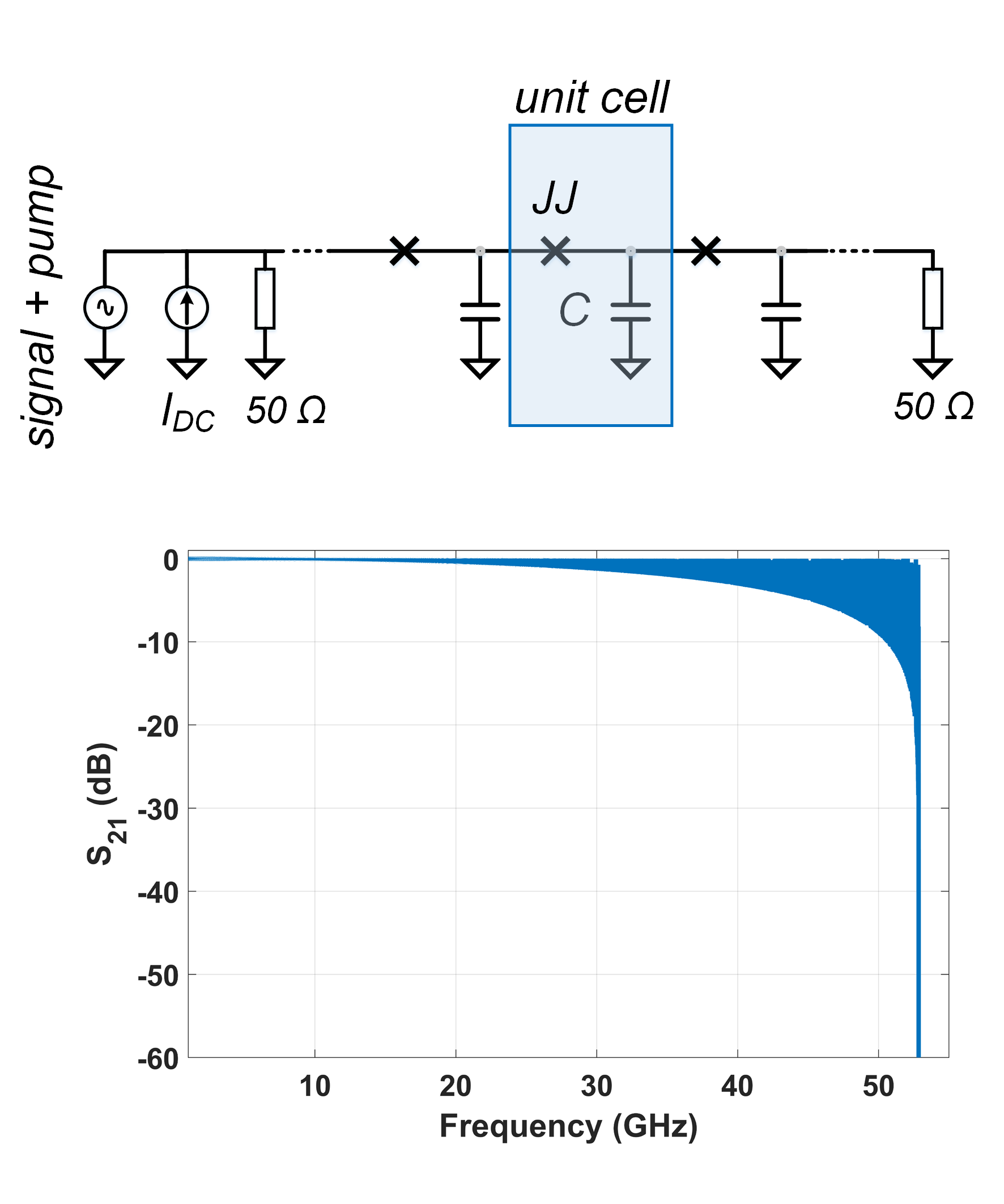}
\caption{(Top) The schematic of a JJ-TWPA. Each unit cell is composed of a series JJ and a parallel capacitor. (Bottom) The simulated magnitude of the transmission $S_{21}$, from $S$-parameter analysis for a TWPA with 2000 unit cells. $I\crit=\unit[1.4]{\micro A}$, $I\dc= I\crit/2$, and $C = \unit[108.6]{fF}$.}
\label{fig_S21JJ}
\end{figure}

The JJ-based traveling-wave parametric amplifier (JJ-TWPA) in this example is composed of $ 1000$ to $2000$ unit cells \textcolor{black}{made of a JJ and a parallel capacitor} as shown in Fig. \ref{fig_S21JJ}. 

\textcolor{black}{In the figure, in addition to the ac source containing the pump and the signal, a dc current is applied to the TWPA input. If the dc current $I_\mathrm{dc} \neq 0 $, the TWPA will support a three-wave mixing (3WM) process during which one photon of the pump frequency generates a signal and an idler photon, $f_\mathrm{p} =f_\mathrm{s}+f_\mathrm{i}$.} The bias current which passes through all series junctions determines the inductance of each JJ. If the critical current of the JJ is $I\crit= \unit[1.4]{\micro A}$, then with $I\dc= \unit[0.7]{\micro A}$, the inductance of each unit cell is $L = \unit[0.2714]{nH}$ \textcolor{black}{according to equation (\ref{eqn_10}).} The TWPA is the discrete implementation of a transmission line, the input impedance of which is determined by the inductance and capacitance of each unit cell, $Z=\sqrt{L/C}$. For a $Z=\unit[50]{\Omega}$ input impedance, the required capacitor is $C = \unit[108.6]{fF}$. The magnitude of the transmission $S_{21}$, for a JJ-TWPA with 2000 JJ's is shown in Fig. \ref{fig_S21JJ}.

The ripples of the response inside the passband result from approximating a transmission line by many discrete unit cells (here 2000 JJ and C). The higher number of transfer function poles leads to a steeper filter response after cut-off; however, this is achieved at the price of increased ripples inside the passband.
The cut-off frequency is found from the dispersion of the circuit, which is $\omega={2}/{\sqrt{LC}}$ \cite{Collin_MW,Hampus2023}. Note that the pump frequency must be lower than the cut-off and within the linear part of the dispersion to simultaneously satisfy phase matching (conservation of momentum of the photon) and conservation of energy. The cut-off frequency of the TWPA can be adjusted by the dc bias current $I\dc$, as it changes the inductance according to equation (\ref{eqn_10}). For $I\dc= \unit[0.7]{\micro A}$, and the aforementioned $C$, the cut-off frequency is 35\,GHz. \\ 
\textcolor{black}{When the bias current of the JJ-TWPA is zero,} the nonlinearity of the TWPA is akin to the third order nonlinear susceptibility coefficient, $\chi^{(3)}$, in nonlinear optical materials \cite{Armstrong1962}. This leads to the 4-wave mixing (4WM) process, where two photons of pump frequency $f_\mathrm{p}$ add up and create a signal and an idler photon (\ie, $f_\mathrm{p}+f_\mathrm{p}=f_\mathrm{s}+f_\mathrm{i}$). Due to this, the TWPA gain spectrum has a mirror symmetry around the pump frequency, for which there is no gain in the spectrum. This can be problematic; if the high-power pump is close to the qubit frequencies, it may cause unwanted excitation of qubits. Putting the signal far away from the pump, on the other hand, yields less gain. 

\textcolor{black}{The dissipation in the TWPA due to the lossy substrate can be modeled by adding a resistor parallel to $C$ in each unit cell of Fig. \ref{fig_S21JJ}. Although the quantum fluctuations cannot be captured with this simulation method, the thermal noise in these strongly nonlinear circuits can be modeled as described in Stephen Maas's book \cite{Maas2005}. The added quantum noise and the quantum efficiency of a JJ-TWPA are calculated using the procedure proposed in \cite{Xaparam2022}, where the X-parameter method in Keysight ADS can be combined with HB analysis. A more general analytical method for noise analysis is also available in \cite{Haider2024}.}

\subsection{Input Impedance with $S$-parameter and HB}
To see the shortcomings of small-signal methods, the input impedance of only one unit cell (JJ and capacitor) using both small-signal (linear $S$-parameter) and large-signal/nonlinear (HB) analyses are compared. In the small-signal analysis, the input impedance is found from the reflection coefficient, $S_{11}$ or the $\Gamma$ coefficient \cite{Collin_MW},
\begin{equation}
\label{eqn_16}
S_{11} = \Gamma=\frac{Z_\text{in}-Z_{0}}{Z_\text{in}+Z_{0}},
\end{equation}

\begin{figure}[!t]
\centering
\includegraphics[width=3.25in]{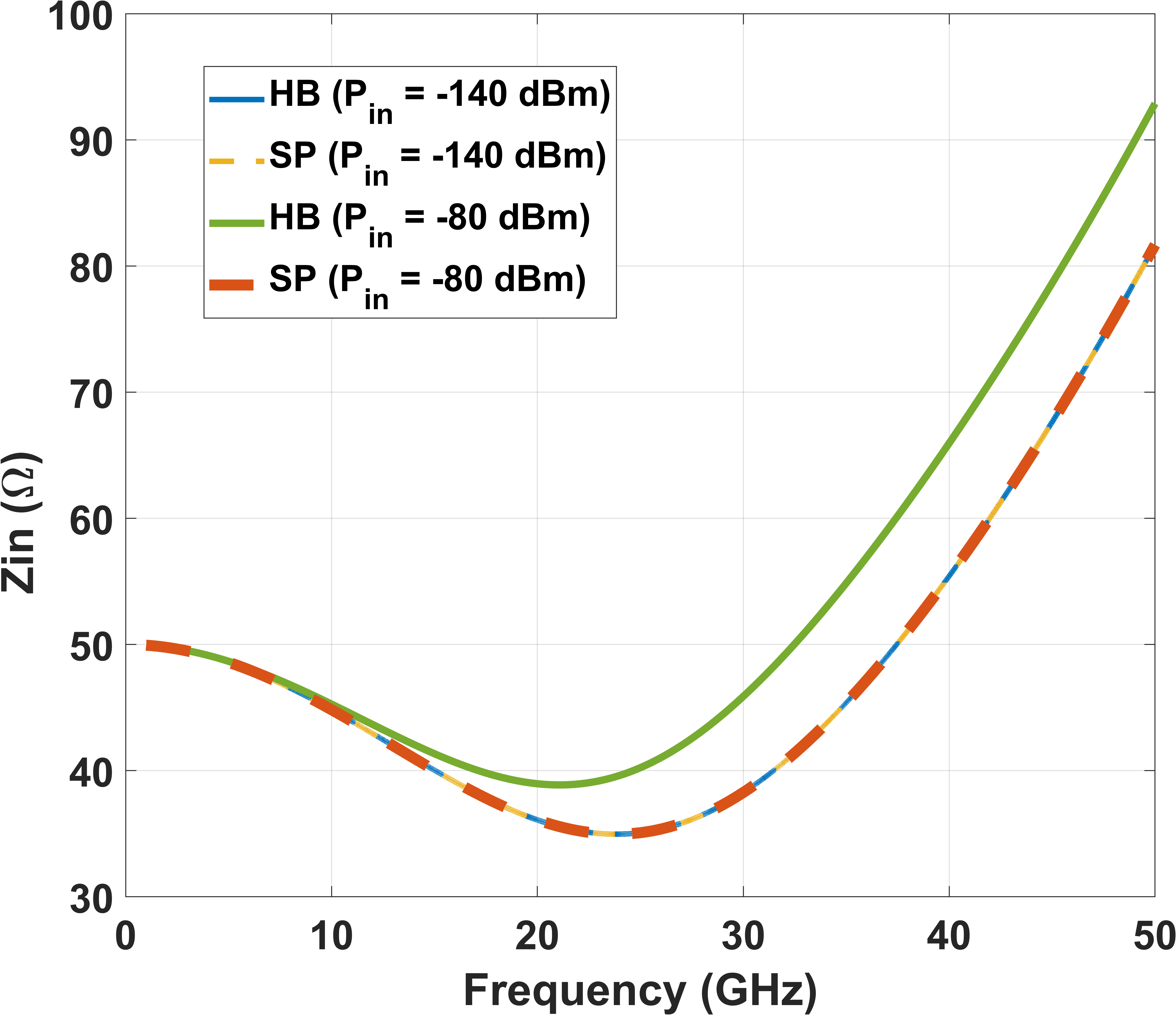}
\caption{The input impedance of a single unit cell (JJ and capacitor) of the JJ-TWPA obtained by HB and S-parameter simulations. The HB method and $S$-parameter analyses give the same result at low power ($P_\text{in} = -\unit[140]{dBm}$) (blue and yellow lines). The deviation appears as the input power is increased ($P_\text{in} = -\unit[80]{dBm}$). \textcolor{black}{The green curve shows the input impedance predicted by HB, while the $S$-parameter result remains} the same despite the power increase. For both cases $I\dc= 0.5I\crit$.}
\label{fig_ZinSvsHB}
\end{figure}

where $Z_{0}=\unit[50]{\Omega}$ is the characteristic impedance of the feed system. In HB analysis, the input impedance is found from the first harmonic of the input voltage divided by the first harmonic of the input current source. For a low power input (\eg, $P_\text{in}=-\unit[140]{dBm}$), the results of the HB method and the $S$-parameter analysis coincide (see dashed lines in Fig. \ref{fig_ZinSvsHB}). However, by increasing the input power to $P_\text{in}=-\unit[80]{dBm}$, the results of the large-signal and small-signal analyses deviate as shown by the solid green line in Fig. \ref{fig_ZinSvsHB}. This example shows why a linear analysis like \(S\)-parameter cannot completely represent the behavior of a strongly nonlinear circuit, and one needs to use HB analysis.

\subsection{Transient-time Analysis}
\begin{figure}[!t]
\centering
\includegraphics[width=3.25in]{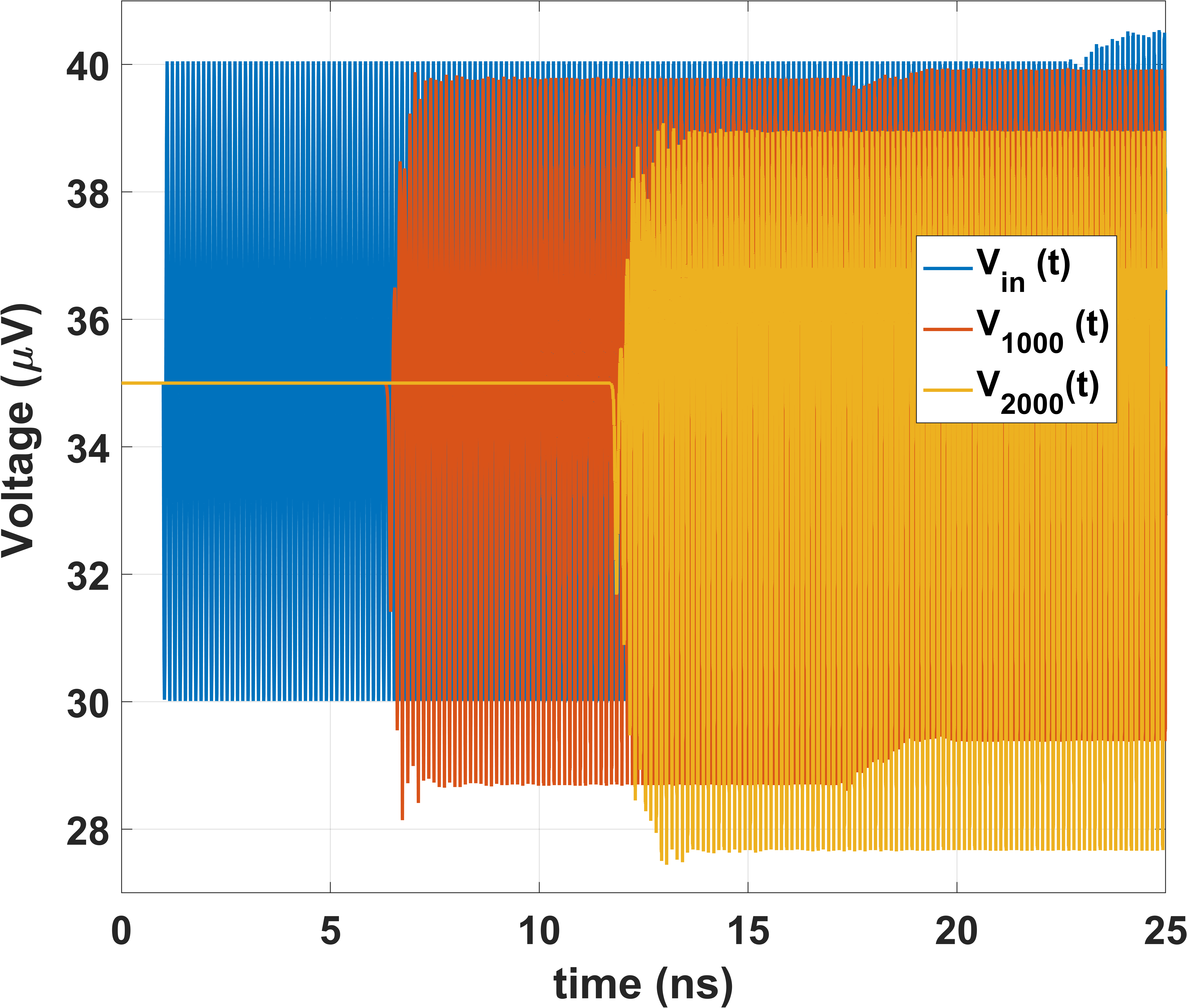}
\caption{Transient-time simulation. The voltage waveforms of the JJ-TWPA at the input, the 1000th, and the 2000th (output) node in response to a single tone at $f\pump=\unit[8]{GHz}$ and $I\pump=\unit[200]{nA}$. The input-to-output delay is $t_d=\unit[11.7]{ns}$. The steps at $\unit[18]{ns}$ and $\unit[23]{ns}$ are due to back and forth reflection of the wave.}
\label{fig_time2KJJ}
\end{figure}

To perform an HB analysis, it is necessary to run a transient (time-domain) simulation first to find out when the circuit reaches the steady-state. Thereafter, the settings in the HB analysis and required parameters (as discussed below) can be selected accordingly. A transient-assisted harmonic balance (TAHB) uses the steady-state time-domain solution of the circuit as an initial guess for HB. The TAHB option can be selected in Keysight ADS along with three important parameters, which are the \textit{Stop Time}, \textit{Max Time Step}, and \textit{Min Time for detecting steady state}. Once the simulation time reaches the \textit{Min Time for detecting steady state}, the HB solver starts working. The transient simulation results of a 2000-JJ TWPA are shown in Fig. \ref{fig_time2KJJ} in which the circuit is fed by a dc current of $\unit[0.7]{\micro A}$ and a single-tone pump of amplitude $I\pump=\unit[200]{nA}$ and frequency $f\pump=\unit[8]{GHz}$. The initial delay of $\unit[1]{ns}$ is intentional for visibility. \textcolor{black}{Fig. \ref{fig_time2KJJ} shows that the wave reaches the first, the $1000$th, and the $2000$th unit cell at $\unit[1]{ns}$, $\unit[7.36]{ns}$, and $\unit[12.7]{ns}$, respectively.} This means it takes $t_\mathrm{d}= \unit[11.7]{ns}$ for the wave to travel through 2000 junctions. This information is useful for determining the three parameters in the settings for a TAHB.\\ 
Additionally, the wave group velocity in the amplifier chain, impedance mismatch, and values of inductance or capacitance per unit length can be extracted from this analysis. For example, knowing that the length of a single JJ is $\unit[15]{\micro m}$, the total length of the TWPA is $l=\unit[2000 \times 15]{\micro m} = \unit[3]{cm}$. The group velocity of the microwave in the TWPA is then $v = \text{distance}/\text{travel~time} = \unit[3]{cm}/\unit[11.7]{ns} = \unit[2.564 \times 10^6]{m/s}$. The group velocity is given by $v = {1}/{\sqrt{L_lC_l}}$ (where subscript $l$ means that the quantities are per unit length). As a test, we can assume that the value of the parallel capacitance in each unit cell is known, which is $C = \unit[108.6]{fF}$. Therefore, $C_l = \unit[108.6]{fF}/\unit[15]{\micro m}$. From this, the inductance value of each JJ is found to be $L = \unit[0.26]{nH}$. This is very close to $L = \unit[0.27]{nH}$, which was calculated based on the JJ nonlinear inductance in equation (\ref{eqn_10}).
In Fig. \ref{fig_time2KJJ}, the reason for the asymmetry in the output waveform around the dc bias is the existence of even harmonics of the injected pump at 8 GHz. \textcolor{black}{As will be explained in the next subsection, when the pump propagates along the TWPA, the even harmonics (in addition to the odd harmonics) are generated and enhanced along the line. This deforms the waveform and reduces its  $50~\%$ duty cycle. As a result, it seems that the average of the wave is decreasing.} The step-like jump in the input wave is the reflection of the output wave back to the input due to a slight impedance mismatch. For HB simulation, the minimum time to reach the steady-state is set to be $t>\unit[40]{ns}$ to assure a good convergence.
  
\subsection{Output Harmonics of a JJ-TWPA}

\begin{figure}[!t]
\centering
\includegraphics[width=3.5in]{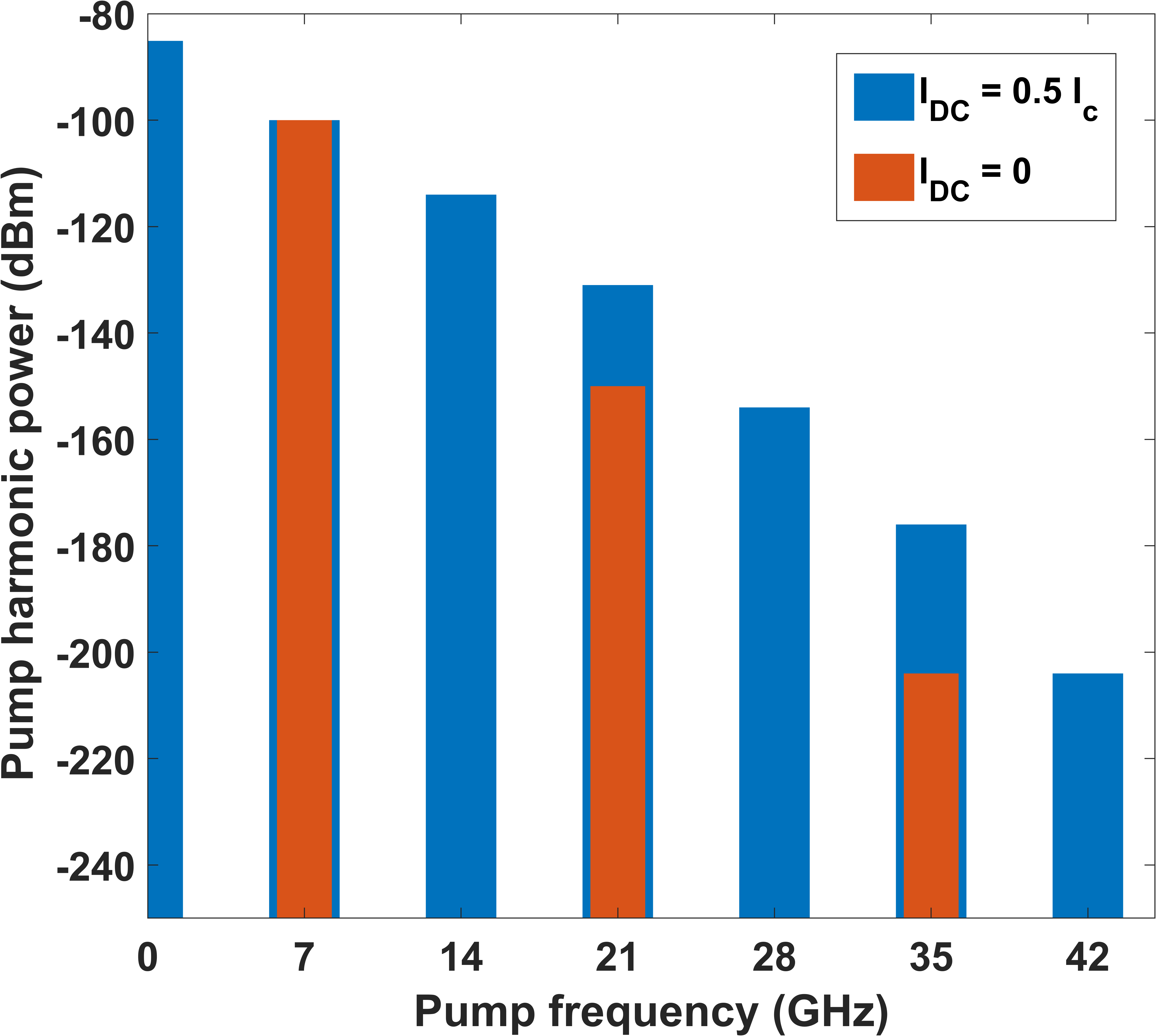}
\caption{Simulated output spectrum of a JJ-TWPA with 2000 unit cells in response to a single tone at \unit[7]{GHz} with $\unit[-100]{dBm}$ power when (red) $I\dc= 0$ and (blue) $I\dc=0.5I\crit$. In the latter case, the even harmonics are also generated.}
\label{harm2KJJ}
\end{figure}

The output spectrum of a JJ-TWPA with 2000 unit cells is shown in Fig. \ref{harm2KJJ} for zero and non-zero dc bias currents and a single input tone at $f = \unit[7]{GHz}$ with $P_\text{in}=-\unit[100]{dBm}$. When there is no dc current, only odd harmonics of the input tone are generated at the output. By applying a non-zero dc bias current ($I\dc=0.5I\crit$), in addition to the odd harmonics, the even harmonics (\eg, 14, 28, and 42\,GHz) are generated at the output.
The reason for this is understood if we consider a single JJ (nonlinear inductor) which is biased by a current source, $I = I\dc +\Tilde{I}\sin(\omega t)$. The voltage which appears on the JJ is found from $v(t)=L_\mathrm{J} \cdot \rd I/\rd t$,
\begin{equation} \label{eqn_Voftharm}
    \begin{aligned}
    v(t)=\frac{L_\mathrm{J0}}{\sqrt{1-\left(\frac{I\dc +\Tilde{I}\sin(\omega t)}{I\crit}\right)^2}}\times\omega\Tilde{I}\cos(\omega t)
    \end{aligned}
\end{equation}
Using the abbreviations as $\alpha ={I\dc}/{I\crit}$ and $\beta={\Tilde{I}}/{I\crit}$, and using the Taylor expansion of equation (\ref{eqn_Voftharm}) we can write
\begin{equation} 
\label{eqn_VofTaylor}
    \begin{array}{c}
    v(t)=L_\mathrm{J0} \omega \beta I\crit \cos(\omega t) \biggl( \frac{1}{\sqrt{1-\alpha^2}} + \frac{\alpha\beta \sin(\omega t)}{(1-\alpha^2)^{1.5}} + \\
    \frac{(2\alpha^2+1)\beta^2 \sin^2(\omega t)}{2(1-\alpha^2)^{2.5}} + \\
    \frac{\alpha(2\alpha^2+3)\beta^3 \sin^3(\omega t)}{2(1-\alpha^2)^{3.5}} + \cdots \biggr).
    \end{array}
\end{equation}

\noindent When $\alpha=0$, $I\dc=0$, and only the frequency components at $\omega$, $3\omega$, $5\omega$, and in general $(2k+1)\omega$ exist in the Fourier spectrum of the JJ voltage ($k$ is an integer number). On the other hand, if $I\dc\neq 0$, then the even harmonics ($2\omega$, $4\omega$, and $2k\omega$ ) are also generated due to frequency mixing. The symmetry (asymmetry) of the steady-state output voltage around the bias value (output dc voltage) also shows if the output spectrum has only odd (or both odd and even) harmonics. \\
\subsection{Gain: Simulation vs. Measurement}
A nonlinear circuit like a TWPA can be considered a mixer in which the signal and pump tones are mixed and different harmonics are generated at the outputs. ADS allows these harmonics to be accessed by calling them by their indices, which is a useful way to obtain the gain spectrum of the amplifier. 

If a nonlinear circuit has two input frequencies (\ie, $f\signal$ and $f\pump$), then at the computed output spectrum, there will be mixed (up-converted) tones at  $nf\signal + mf\pump$, where $n$ and $m$ are integer numbers. To address each component of the spectrum at an arbitrary node, the command mix(node~name,\{$n,m$\}) is used. Note that in the \textit{Fundamental Frequencies} list of the HB settings, $f\signal$ and $f\pump$ should be listed as the first and the second tones to correspond with $n$ and $m$, respectively \cite{Keysight}. To access the power of the signal component at the output, the power of the tone, which corresponds to $\{n,m\}  = \{1,0\}$, is plotted. Likewise, to see how the amplitude of the pump harmonics changes as a function of the signal frequency or the length of the amplifier, the mixed component $\{0,m\}$ is chosen. To access the idler frequency, $\{n,m\}  = \{-1,1\}$ is chosen, because the idler frequency is obtained from $f\idler = -f\signal + f\pump$.
The power gain in dB is then found by subtracting the output power from the input power, both at $\{1,0\}$ and expressed in dBm. The signal frequency, in this case, is chosen as a sweep parameter to plot the gain versus $f\signal$. 

\begin{figure}[!t]
\centering
\includegraphics[width=3.5in]{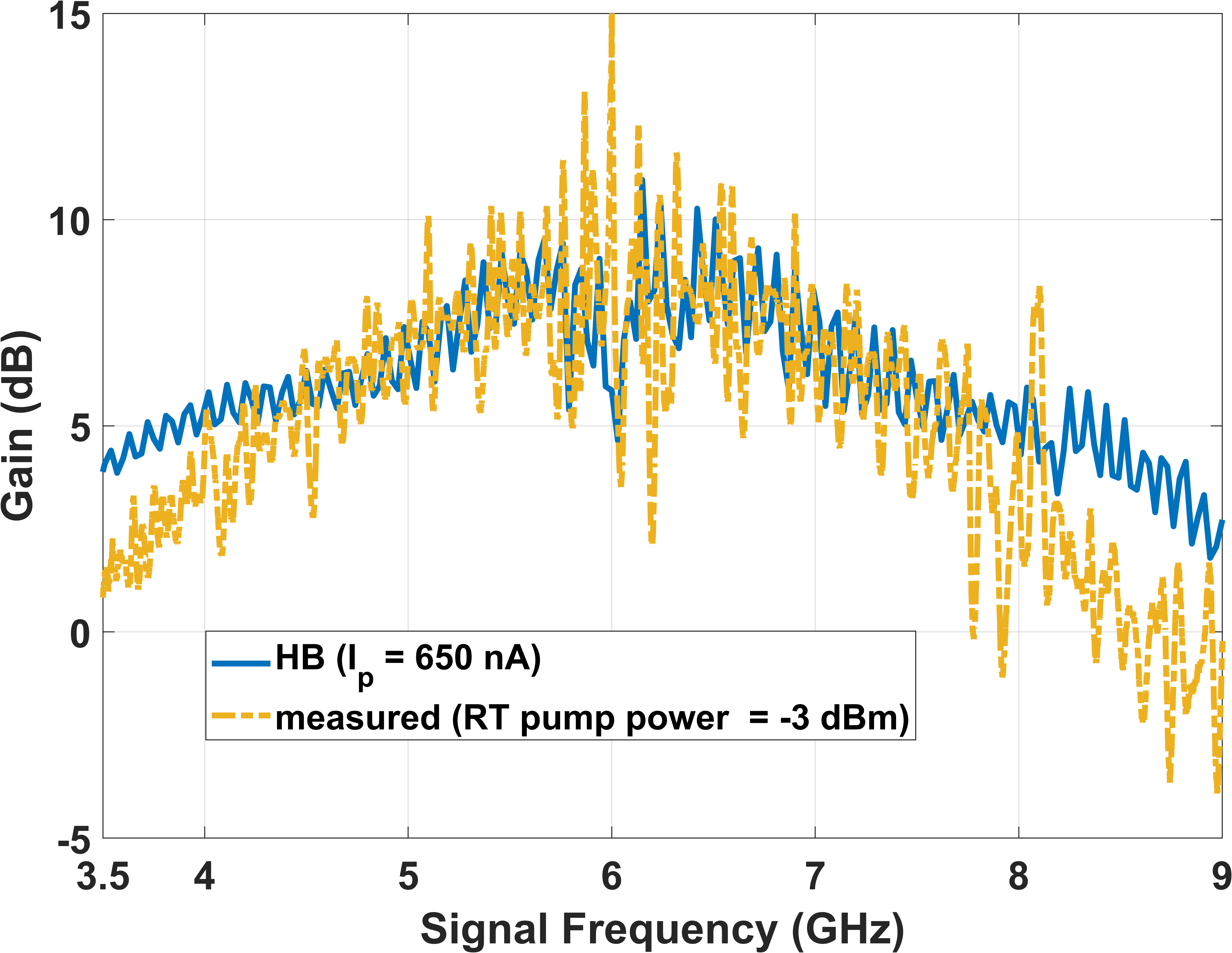}
\caption{The gain spectrum of a JJ-TWPA with 1000 unit cells obtained by the HB method and measurement. The pump power reported for the experimental data is measured at room temperature ($P_\mathrm{RT}$) at the signal generator output. Note that the gain is due to the 4WM process as $I\dc= 0$, and it is symmetric around the pump frequency.}
\label{fig_gain1Kjj}
\end{figure}

In Fig. \ref{fig_gain1Kjj}, the simulated and measured gain spectra of a JJ-TWPA with 1000 unit cells are shown. This amplifier has JJs with critical current $I\crit=\unit[1.318]{\micro A}$ and operates with 4-wave mixing, since $I_\mathrm{dc} = 0$. The value of the capacitor in each unit cell is $C = \unit[93]{fF}$ to obtain $Z_\text{in}=\unit[50]{\Omega}$. The pump is injected at $f\pump=\unit[6.0102]{GHz}$ with an amplitude of $I\pump=I\crit/2$. The pump frequency in the simulation is slightly different from $\unit[6]{GHz}$, to avoid the generation of mixing products that land at the same frequency. As can be seen in Fig. \ref{fig_gain1Kjj}, the simulated and measured values are close, especially within the band of interest $\unit[4-8]{GHz}$ where the gain maximum reaches $\unit[10]{dB}$. The discrepancy between measurement and simulation occurs below $\unit[4]{GHz}$ and after $\unit[8]{GHz}$, where the measured gain is lower than the simulation. This is attributed to the bandpass filters and post-amplifiers along the measurement line which are centered on the \unit[4 - 8]{GHz} range, and out of this range, there is attenuation. The measurement setup of the TWPA is shown in Appendix B and discussed in detail in \cite{Anita2023} \textcolor{black}{where the added noise of the TWPA is also extracted.} 

\section{SNAIL-based TWPA}
\begin{figure}[!t]
\centering
\includegraphics[width=3.5in]{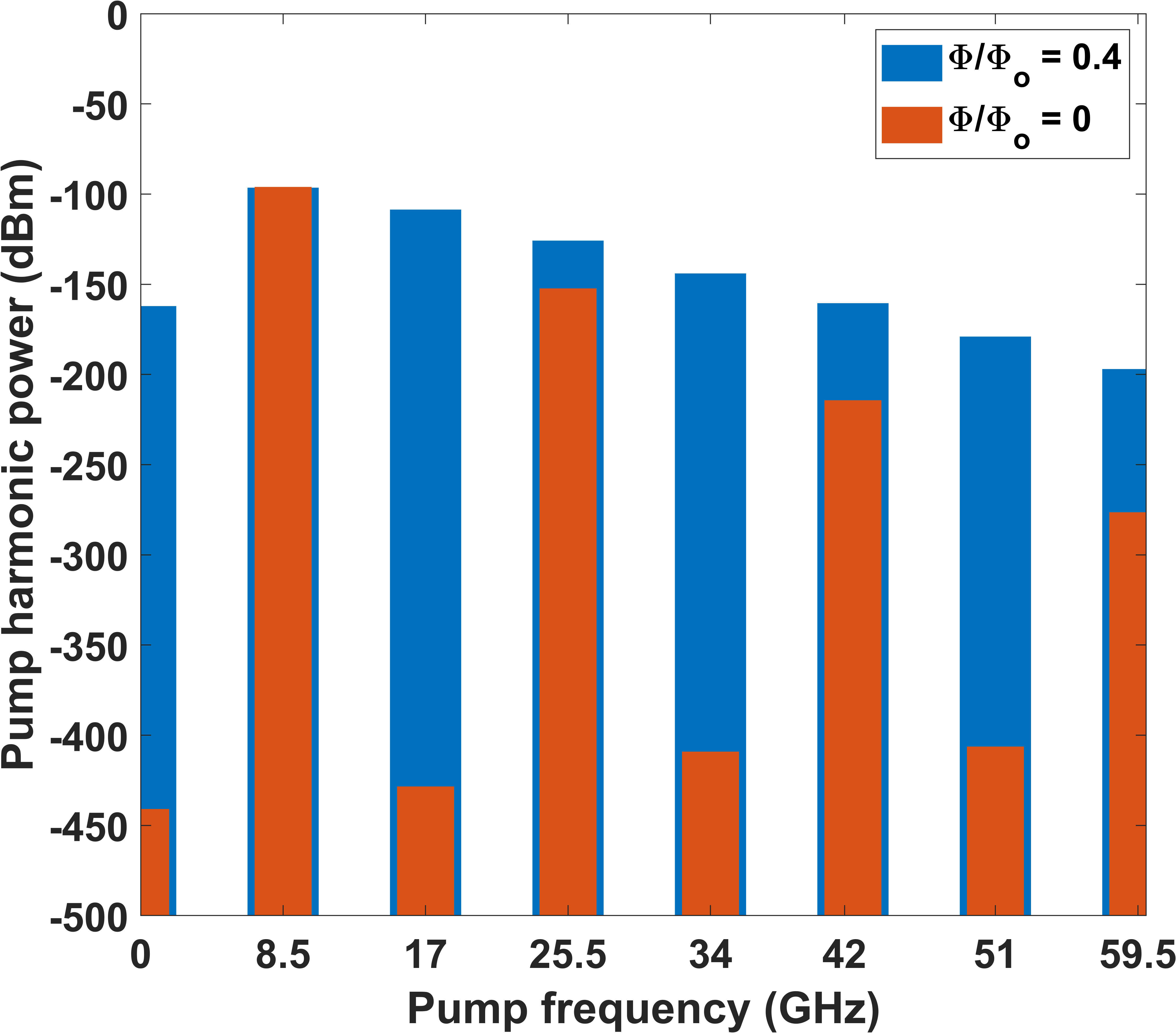}
\caption{Simulated output spectrum of a SNAIL-TWPA with 440 unit cells in response to a single tone at \unit[8.5]{GHz} with $I\pump=\unit[100]{nA}$ when (red) $\Phi/\Phi_0 = 0$ and (blue) $\Phi/\Phi_0 = 0.4$. In the latter case, even harmonics are also created.}
\label{KerrEffect}
\end{figure}

\begin{figure}[!t]
\centering
\includegraphics[width=3.5in]{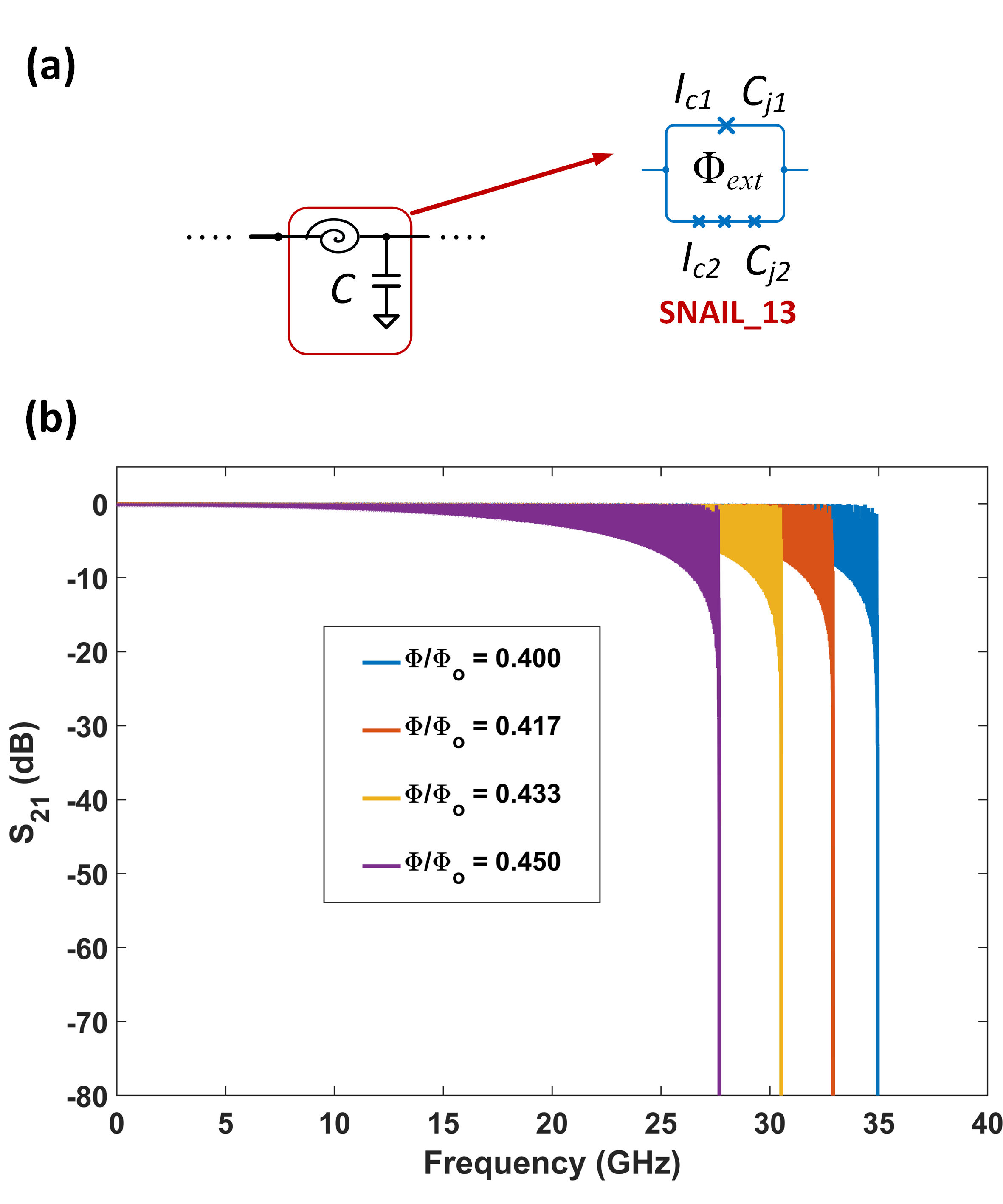}
\caption{(a) The schematic of a SNAIL-TWPA unit cell. It contains a SNAIL-13 and a parallel capacitor $C$. (b) The simulated magnitude of the transmission $S_{21}$ for different flux biases from 0.4 to 0.45 of a flux quantum.}
\label{fig_SnailS21vsF}
\end{figure}

TWPAs based on SNAILs promise higher gain with a smaller number of unit cells \cite{Frattini2017,Anita2023,Hampus2023}. This is because a SNAIL at a non-zero magnetic flux bias enhances the three-wave mixing process. Moreover, the pump frequency is always above and out of the gain spectrum, which mitigates the unwanted excitation of qubits due to a strong pump. 
Designing SNAIL-TWPAs starts with finding the optimum flux bias. First, the total potential energy of the SNAIL in equation (\ref{eqn_Etot}) is expanded as a power series of $\phi$ around the minimum of the potential. The terms that are proportional to $\phi^3$ and $\phi^4$ are responsible for the 3WM and 4WM processes, respectively. By adjusting the flux bias, it is possible to enhance the 3WM and reduce the 4WM terms in the potential energy\cite{Anita2023,Hampus2023,Zorin_flux2019,Frattini2017,Ranadive2022,Haider2024}. 
\textcolor{black}{Before moving to S-parameter analysis and the gain spectrum, it is instructive to examine the effect of flux bias on the 4WM term in SNAIL-based TWPAs. In Fig. \ref{KerrEffect}, we show the HB simulation results of a 440-unit cell SNAIL-TWPA with a single input pump tone at $f\pump = \unit[8.5]{GHz}$ and $I\pump = \unit[100]{nA}$. Each unit cell contains one SNAIL-13 and a capacitor to ground. The schematics of the amplifier and each unit cell are shown in Fig. \ref{fig_SnailS21vsF}(a). The Josephson junctions are designed such that $I\crit[1]=\unit[3]{\micro A}$, $C_{j1}=\unit[8.2]{fF}$, and $\alpha=1/3.75$. The capacitor in each unit cell is $C=\unit[150]{fF}$ to make the input impedance $\unit[50]{\Omega}$. For this structure the optimum flux bias to suppress 4WM is $F = \Phi/\Phi_0 = 0.4$. At this optimum point all harmonics of the pump exist at the output spectrum, which shows that the 3-wave mixing process is dominant (see Fig. \ref{KerrEffect} for $F = \Phi/\Phi_0 = 0.4$). When the flux bias is zero, the 4-wave mixing process is dominant. This manifests itself at the output spectrum which has only odd harmonics of the pump ($F =\Phi/\Phi_0 = 0$) as shown in Fig. \ref{KerrEffect}.}

Before running the HB analysis, an $S$-parameter analysis is performed to check the cut-off frequency of the amplifier and its dependence on the applied dc flux. Fig. \ref{fig_SnailS21vsF}(b) shows that when the external flux bias is changed from $0.4\Phi_0$ to $0.45\Phi_0$, the bandwidth of the circuit is reduced due to increased inductance of the SNAILs. Note that for each flux value, the capacitance is readjusted to keep the input impedance of the TWPA the same.

\subsection{Gain: Simulation vs. Measurement}
\begin{figure}[!t]
\centering
\includegraphics[width=3.5in]{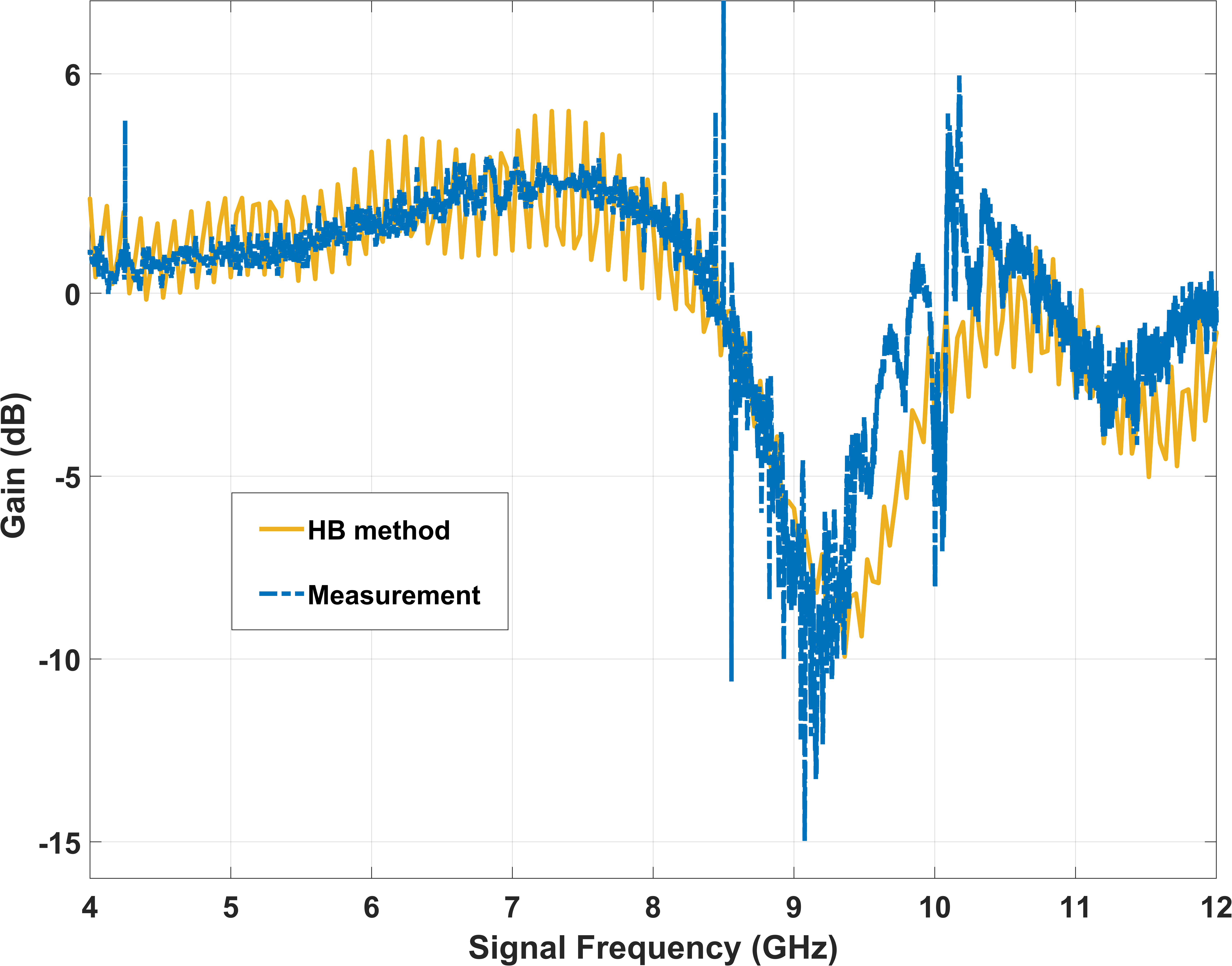}
\caption{The gain spectrum of a SNAIL-TWPA with 440 unit cells. \textcolor{black}{For unit cell parameters see Fig. \ref{fig_SnailS21vsF}.} Blue and yellow traces are the measurement and HB simulation, respectively. In the simulation, the signal current is 1/1000th of the pump. In the experiment, the input pump power is $-\unit[97]{dBm}$. The pump frequency is $\unit[8.5]{GHz}$.}
\label{fig_ExpSimSNAIL}
\end{figure}

A time-domain (transient) analysis is performed, similar to the one presented for JJ-TWPA. From this step, the important parameters for TAHB are chosen to obtain good convergence of the HB analysis. Thereafter the gain and harmonic contents of the output in the presence of a pump are investigated using the HB method. 
The signal amplitude is less than one-thousandth of the pump, $I\signal < I\pump/1000$. The amplitude of the input pump current is $I\pump=\unit[100]{nA}$ and the frequency is $f\pump=\unit[8.5]{GHz}$. The magnetic flux bias is $0.4\,\Phi_0$. Fig. \ref{fig_ExpSimSNAIL} shows the gain spectrum of the SNAIL-TWPA, which agrees well with the one measured at approximately the same input pump power and the same pump frequency, $f\pump=\unit[8.5]{GHz}$. 

\textcolor{black}{In Fig. \ref{fig_ExpSimSNAIL}, the oscillation in the gain spectrum is due to multiple reflections of the wave at the input and output ports of the TWPA which emanate from impedance mismatch. It is possible to calculate the group velocity of the wave in the TWPA from the period of these ripples. The gain ripples were investigated by modeling the TWPA as a Fabry-Perot cavity, in which both imperfectly matched ends of the TWPA are modeled as semi-reflective mirrors. The gain ripple was then reproduced successfully in a spatially modulated SQUID TWPA (4WM) \cite{Planat2020} and a SNAIL-13 TWPA (3WM) \cite{Haider2023}. However, the impedance mismatch (and gain ripples) are dependent on the pump power and the impedance of each unit cell is perturbed by the generation of the idler due to the strong pump. The model and methods to mitigate the gain ripples to obtain a flat gain spectrum using hybrid couplers and diplexers are discussed in \cite{Hampus2024}.}

In the yellow line (HB analysis) in Fig. \ref{fig_ExpSimSNAIL}, the frequency difference between two consecutive ringings is about $\Delta f=\unit[160]{MHz}$. This corresponds to a time delay of $t_d = \unit[6.25]{ns}$. Transient simulation shows that this is indeed one period during which the wave traverses the length of the TWPA back and forth. 
At frequencies below the pump frequency there is a broader range of gain (\eg, from $f\signal=\unit[5 - 8]{GHz}$). When the signal frequency reaches half of the pump, $f\signal=\unit[4.25]{GHz}$, an idler at the same frequency is generated. This manifests itself by a sharp peak at $\unit[4.25]{GHz}$ in Fig. \ref{fig_ExpSimSNAIL}. Also note that, as opposed to that in Fig. \ref{fig_gain1Kjj}, a broader band-pass measurement line was used and as a result, there is less difference between the simulated and measured gains. \\
For this 440-unit cell SNAIL-TWPA, the simulation of the gain spectrum for one input pump power, one input signal power, and 210 different input signal frequencies takes about 2.25 hours. Note that, depending on the power of the pump, the time for TAHB convergence varies from 15 min to 25 min.
As expected from the 3WM process, the pump sits at the high end of the gain spectrum, $f\pump=\unit[8.5]{GHz}$. The exchange of power between the harmonics of the pump is a mechanism that causes a reduction of gain in the 3WM process. In the next section, we show where this exchange mechanism emanates from. 

\section{Power exchange between pump harmonics}
Due to the strong nonlinearity and high power of the pump tone, higher harmonics of the pump are generated inside the bandwidth of the TWPA. The inter-mixing and power exchange between these harmonics cause a reduction of gain \cite{Anita2023,Hampus2023,Dixon2020,Peng2022}. This section briefly reviews the coupled-mode equations for power conversion between pump harmonics, and compares their predictions with HB simulations. We show agreement between HB analysis and coupled-mode theory in capturing this effect. Proposals to stop the power exchange between the harmonics of the pump and enhance the gain in the 3WM regime are discussed in \cite{Anita2023,Hampus2023,Zorin_pmatch2021}.

The coupled-mode equations describe the evolution of the amplitude of each pump tone as they propagate along the length of the amplifier. The study starts with writing the discrete form of the wave equation for phase, $\phi$, which resembles the wave equation of a transmission line, except for the extra nonlinear terms due to the dependence of the group velocity on the inductance via equations (\ref{eqn_10}) and (\ref{eqn_15}), for JJ and SNAIL, respectively. Using the trial solution as a linear combination of pump tones, $\phi(x,t)=\sum_{m=1}^{M}A_me^{i(k_mx-\omega_mt)}$, the following re-scaled coupled-mode equation is obtained. This equation couples each harmonic $m$ to the rest of the harmonics above and below its frequency within the range of $M$ harmonics \cite{Hampus2023}.     
\begin{equation}
\begin{aligned}
    \frac{da_m}{d\xi} = m \Bigg( \sum_{n=m+1}^M &a_n a_{n-m}^* \mathrm{e}^{i\mu\xi d_{n,m}} 
    - \frac12 \sum_{n=1}^{m-1} &a_n a_{m-n} \mathrm{e}^{-i\mu\xi d_{n,m}} \Bigg),
\end{aligned}
\label{eqn_coupled}
\end{equation}

\noindent where \(\xi\) is the effective (normalized) length and is given by
\begin{equation}
    \xi(x) = \frac{c_3\omega_1^2A_1(0)x}{4a\omega_0^2}, 
\label{eqn_xi}
\end{equation}
The normalized amplitude of the wave $a_m$ is
\begin{equation}
a_m = m\frac{A_m(x)}{A_1(0)}. 
\label{eqn_am}
\end{equation}
where \(A_1(0)\) is the amplitude of the first pump harmonic at the input port. The effective phase mismatch is
\begin{equation}
    \mu = \frac{k_2-2k_1}{c_3\omega_1^2A_1(0)/(4a\omega_0^2)},
    \label{eqn_mu}
\end{equation}
and $d_{m,n} = \frac12 mn(m-n)$ is a numerical factor, \(c_3\) is the 3WM coefficient in the power series expansion of potential energy versus $\phi$, \(\omega_1\) is the frequency of the first harmonic, \(a\) is the physical length of a unit cell, \(x\) is the physical length variable, \(k_1,k_2\) are the wave numbers for the first and second harmonic, and \(\omega_0\) is the resonance frequency of each unit cell, $1/\sqrt{LC}$.

\begin{figure}[ht]
    \centering
    \includegraphics[width=3.5in]{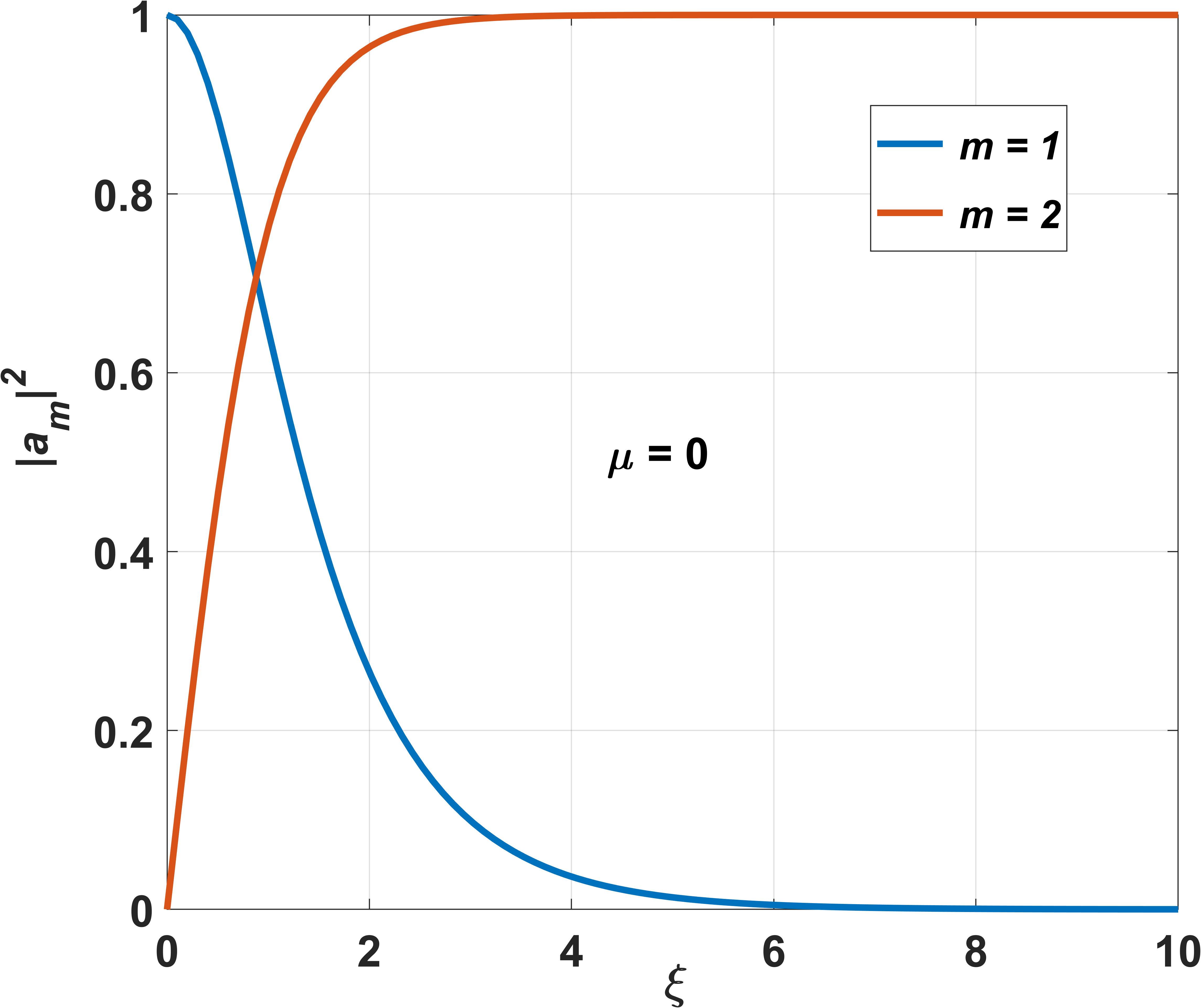}
    \captionsetup{justification=centering}
    \caption{The power exchange between the first and the second harmonic of the pump obtained by coupled-mode equations. The power of the second harmonic $a_{2}$ reaches that of the first harmonic $a_{1}$ after the frequency conversion distance $\xi=1$. Note that phase matching is assumed (\ie, $\Delta k=0$ or $\mu=0$).}
    \label{pumpdep_simple}
\end{figure}

The simplest case is when we assume $M=2$ and perfect phase matching ($\mu=0$), for which equation (\ref{eqn_coupled}) has an exact analytic solution. 
\textcolor{black}{The solutions are $a_1(\xi) = \mathrm{sech}(\xi)$ and $a_2(\xi)=\tanh(\xi)$ \cite{Armstrong1962}.
Fig. \ref{pumpdep_simple}} shows how the power of the main tone of the pump is converted to the second harmonic as the wave propagates inside the amplifier. This process, second harmonic generation (SHG), has also been observed in nonlinear optical materials. The cases for $M=2$ and $M=3$ (third harmonic generation) were studied by \cite{Armstrong1962}, which corroborates what is calculated here for a TWPA. 

In Fig. \ref{pumpdep_simple} the power exchange between $a_1$ and $a_2$ takes place after a certain period. This period is called the \textit{frequency conversion distance} (FCD), and it is where the power of the generated second harmonic becomes as strong as the first harmonic. Note that throughout the exchange process, the power conservation is satisfied (\eg, $\left|a_1\right|^2 + \left|a_2\right|^2=1$). When the pump strength is large in relation to the phase mismatch (\ie, when $\mu$ in equation (\ref{eqn_mu}) is small), more harmonics are generated which need to be included in the coupled-mode equation (\ref{eqn_coupled}) by increasing $M$. Furthermore, the assumption of phase matching is not always valid, and that leads to oscillations of power along the length of the amplifier. For the simulations in this section, we use a SNAIL-13 with $I\crit[1]=\unit[2.53]{\micro A}$, $I\crit[2]=\unit[1.26]{\micro A}$, $C_\text{SNAIL}=\unit[25]{fF}$ and a ground capacitance of $C_0=\unit[159]{fF}$ to make the input impedance approximately $\unit[50]{\Omega}$. We use the flux bias of \(\Phi_\mathrm{ext} = 0.45\,\Phi_0\), which eliminates 4WM while giving the 3WM coefficient of \(c_3=1.11\).

\begin{figure}[h]
    \centering
    \includegraphics[width=3.5in]{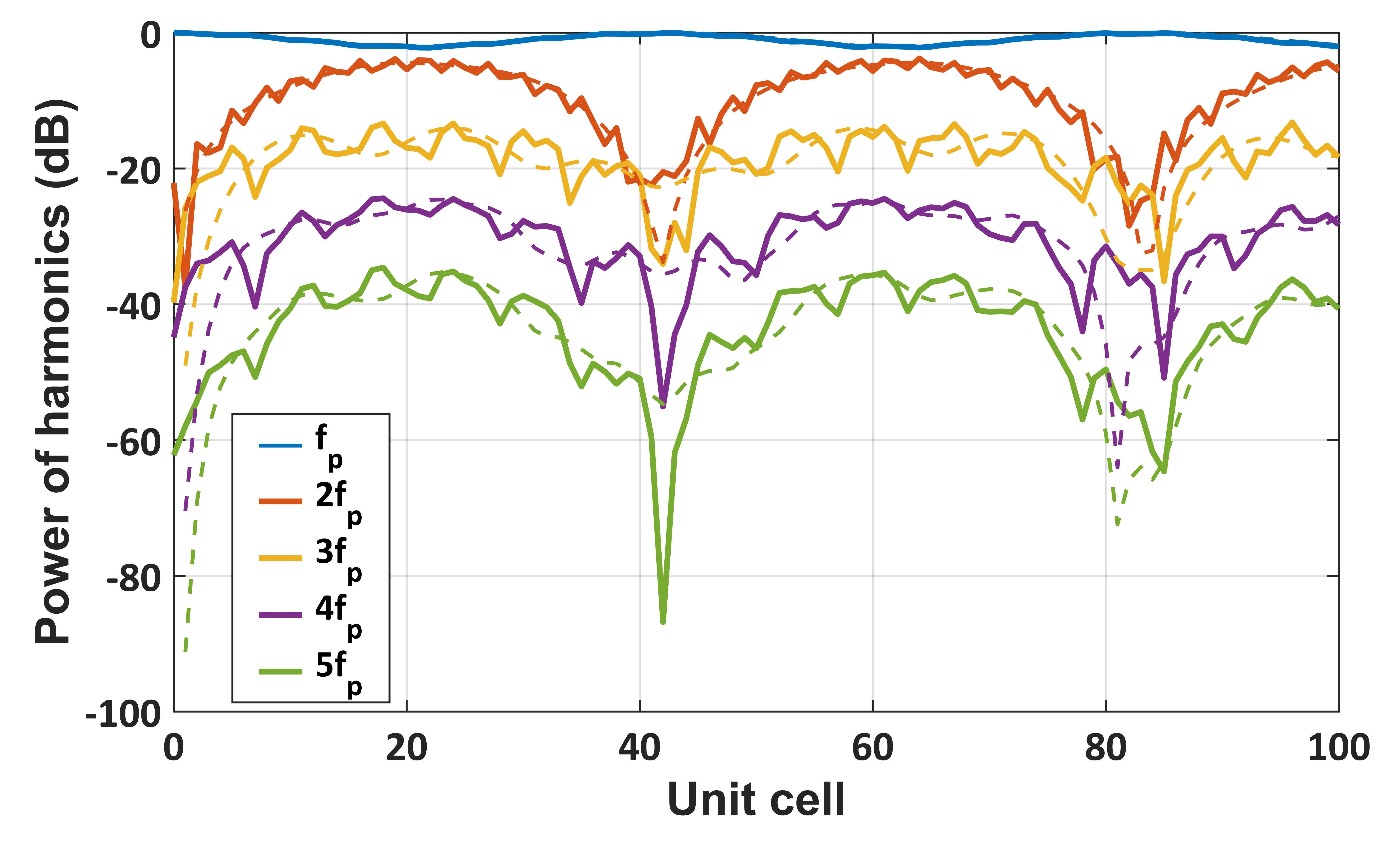}
    \captionsetup{justification=centering}
    \caption{Power exchange between five harmonics of the input pump tone in a 100-unit cell SNAIL-TWPA. The main tone frequency, current, and flux bias are $f\pump = \unit[10]{GHz}$, $I\pump = \unit[400]{nA}$, and $F= 0.45$, respectively. Solid lines are  HB simulations, and dashed lines are the solutions of the coupled-mode equations [equation (\ref{eqn_coupled})].}
    \label{pump_osc}
\end{figure}

In Fig. \ref{pump_osc}, the exchange of power between five harmonics of the pump tone is shown as a function of the amplifier length. As can be observed, there is a good agreement between HB (solid lines) and coupled-mode numerical solution (dashed lines). However, HB is able to show more detailed features and oscillations. 
This means that, as opposed to the HB simulation, which inherently shows the oscillation and power exchange of all harmonics along the length of the amplifier, the coupled-model theory catches these features only if $M$ is increased.
The pump current and its first tone frequency are $I\pump = \unit[400]{nA}$ and $f\pump = \unit[10]{GHz}$, respectively. As can be seen in Fig. \ref{pump_osc}, the power in the second harmonic oscillates approximately every 40 unit cells, which is $\approx \unit[40\times15]{\micro m}$ long. Increasing the pump power and/or the pump main tone frequency leads to a faster oscillation of power exchange between the pump harmonics, as predicted by equation (\ref{eqn_xi}). 

This study suggests that by adding a stop-band in the TWPA (\ie, a band gap in the dispersion of the TWPA), it is possible to suppress the second harmonic. As a result, the coupling between the higher harmonics and the main tone is cut. Having a dispersion-engineered band gap on the second harmonic of the pump indeed led to the gain improvement \cite{Macklin2015,Anita2023,Peng2022} as it quenched the power oscillation mechanism in Fig. \ref{pump_osc}. 
The same process of power conversion in Fig. \ref{pump_osc} was observed for a JJ-TWPA with 500 unit cells using a combination of time-domain and Fourier analysis \cite{Dixon2020} where the JJ was modeled in WRspice. The same behavior was also observed in kinetic-inductance TWPAs \cite{Shan2016}.\\ 

\begin{figure}[h]
    \centering
    \includegraphics[width=3.5in]{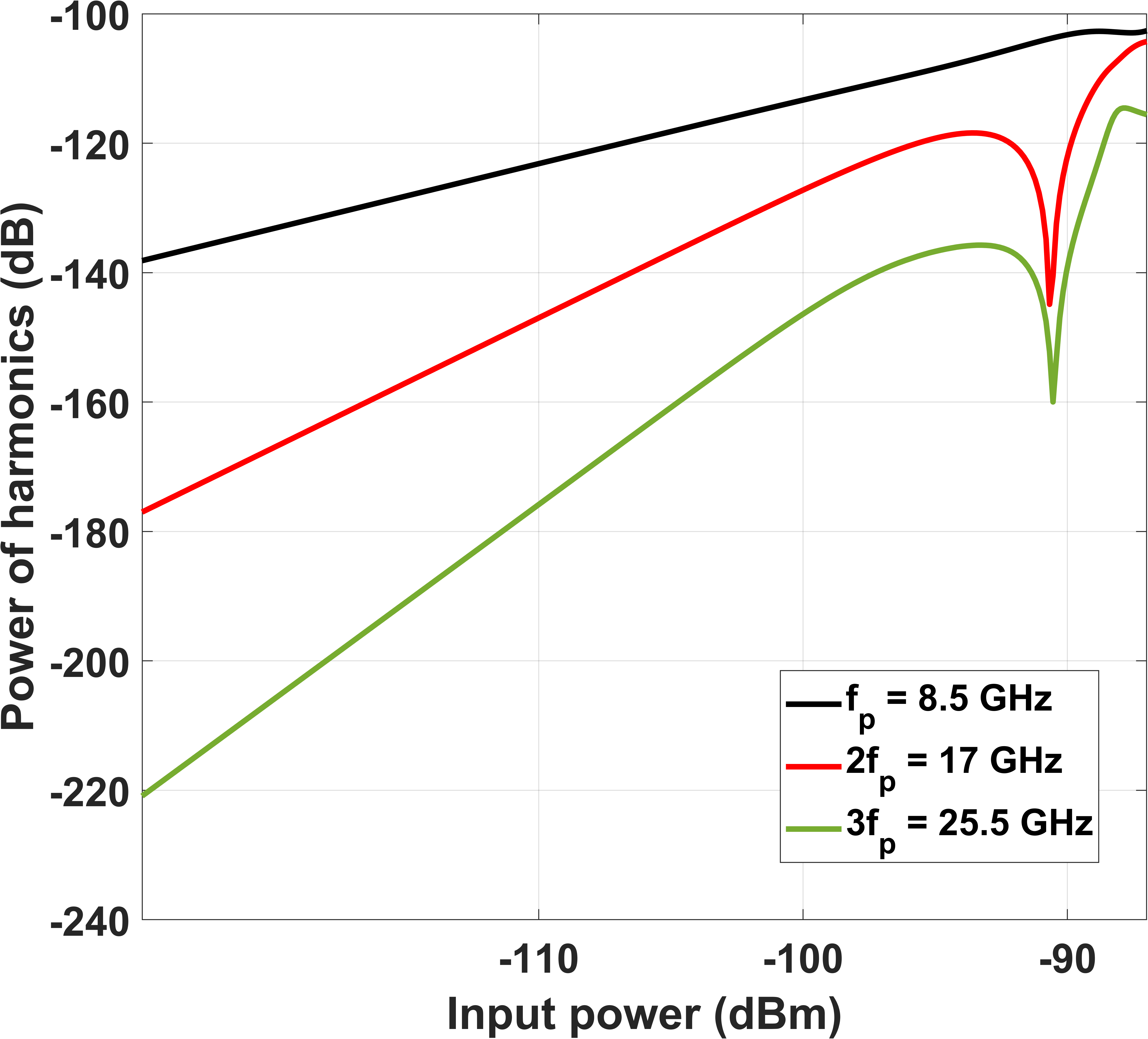}
    \captionsetup{justification=centering}
    \caption{Simulated power of three pump harmonics at the output versus the power of the main pump tone at the input of a 440-unit cell SNAIL-TWPA. The slopes have a ratio of 1:2:3 at low power. The main tone frequency is $f\pump= \unit[8.5]{GHz}$, and the flux bias is $F= 0.4$.}
    \label{harm123_snail}
\end{figure}

A similar behavior is shown in Fig. \ref{pump_osc} and Fig. \ref{harm123_snail} for a JJ-TWPA with 2000 unit cells (with 4WM), which this shows the computational power the HB method offers. \textcolor{black}{It was demonstrated previously that periodically (and capacitively) loading each unit cell with a resonator, a technique called resonant phase matching (RPM), could lead to better phase-matching and gain enhancement in TWPAs based on a 4WM process \cite{OBrien2014}. A larger gain and a broader bandwidth in a full quantum model without and with losses were reported for 4WM TWPAs using RPM in \cite{vanderReep2019,Yuan2023}. In \cite{Planat2020}, the size of the SQUIDs in the TWPA is modulated to create a bandgap (notch) in $S_{21}$.}

Plotting the power of the pump harmonics versus the power of the main input tone yields useful information. In the HB analysis, if there is only one input tone, the power of the output harmonic at each node of the amplifier is accessible using the mix ADS command as dBm(mix(node~name, $\mathbb{N}$)), where $\mathbb{N}$ is the harmonic number. The power of each harmonic at each node can also be plotted versus a given sweep parameter. In this case the useful command is plot\_vs, where the number of the harmonics, $\mathbb{N}$ is used as, plot\_vs(dB(node~name[::, $\mathbb{N}$]), Sweep~parameter). With this, the power of the first three pump harmonics at the output of the SNAIL-TWPA (design shown in Fig. \ref{fig_SnailS21vsF}) is plotted versus the input power in Fig. \ref{harm123_snail}. In the low power regime (below $-\unit[110]{dBm}$), the power of three harmonics is growing linearly, and the ratio of their slopes is $1:2:3$. This is expected from a nonlinear transfer function like $y(x)=y_o+ax+bx^2+cx^3+...$, where $y$ is the output and $x$ is the input defined as $x=A\sin(\omega t+\beta)$. This transfer function is implemented by the nonlinear inductance of a JJ or a SNAIL. 

Note that at high pump power (\ie, $P_{in}>-100~dBm$), the compression of the main input tone appears as the power exchange between all harmonics takes place. The dip in the harmonics power is the beginning of the oscillatory behavior of the pump harmonics versus input power. This is because the effective phase mismatch [in equation (\ref{eqn_mu})] is also a function of input pump power $A_1(0)$
\begin{equation}
    \mu \propto \frac{1}{\omega_1^2A_1(0)}.
\label{eqn_muprop}
\end{equation}
\\
As the input pump power increases, $\mu$ changes and this results in the oscillation of the amplitudes of pump harmonics as a function of power. Recall that the change of each harmonic amplitude along the length of the TWPA is governed by equation (\ref{eqn_coupled}) which in its simplified form looks like the following (for simplicity assume $M=2$)
\begin{equation}
    \begin{aligned}
    \frac{\mathrm{d}a_1}{\mathrm{d}\xi} = \text{const.}\cdot a_2\cdot a_1^*\sin(\mu\xi)\\
    \frac{\mathrm{d}a_2}{\mathrm{d}\xi} = -\text{const.}\cdot a_1^2 \sin(\mu\xi). \\
\end{aligned}
\label{eqn_simplewave}
\end{equation}
\\
This means the pump harmonics oscillate along the length of the TWPA at a constant pump frequency and amplitude. On the other hand, at a fixed point along the TWPA (\eg, at the load section), they oscillate due to a change in pump power (recall $\mu \propto 1/A_1(0)$).

\section{Conclusion}
The harmonic balance method provides a wealth of information about traveling-wave parametric amplifiers based on JJs and SNAILs. First, we showed how JJs and SNAILs can be modeled mathematically using symbolically defined devices. The nonlinearity of these devices is included in its entirety in the model, and there is no need for any approximations or extra analytical/numerical analysis. Second, from the HB analysis, the parameters like gain spectrum, harmonic content of the amplifier output or any intermediate nodes, and the mechanism of power conversion between the pump harmonics were presented. \textcolor{black}{The combination of this method and the X-parameter simulations \cite{Xaparam2022} or the full quantum model in \cite{Yuan2023} can be used to extract the noise behavior of TWPAs.} The comparison and close match of the HB simulation results with the experimental measurements and coupled-mode theory shows the reliability of equation-based modeling and HB analysis in addressing the nonlinear physics of these amplifiers.

\section*{Acknowledgments}
This research was funded by the Knut and Alice Wallenberg (KAW) Foundation through the Wallenberg Center for Quantum Technology (WACQT). The authors acknowledge
the use of the Nanofabrication Laboratory (NFL) at Chalmers University of Technology. 

\section*{Appendix A}
\section*{Harmonic Balance Method}
\begin{figure}[!t]
\centering
\includegraphics[width=3.5in]{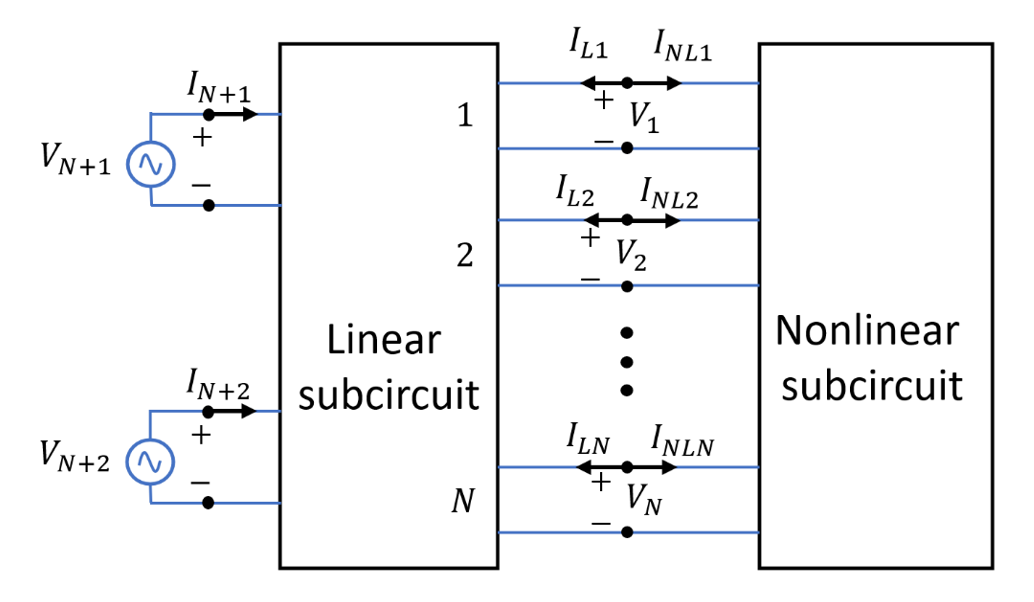}
\caption{Partitioning a circuit into linear and nonlinear sub-circuits. The HB algorithm finds the unknown vector of the common port voltages.}
\label{fig_1HB}
\end{figure}

In the HB method, splitting a circuit into a linear part and a nonlinear part is done similar to Fig. \ref{fig_1HB} where the voltages of common $N$ ports between linear and nonlinear sections are shown by $V_1, V_2, ...,V_N$. Also, the circuit has a few excitation ports which are connected to the linear sub-circuit as ports $N+1$, $N+2$, $\cdots$. Without loss of generality, we assume the split circuit has two common ports between the linear and nonlinear sub-circuits (\ie, $N = 2$) and there is one excitation port from which a single tone at frequency $\omega\pump$ drives the circuit. 
The currents of the linear and nonlinear parts are shown by subscripts L and NL, respectively. The common voltages of the ports are shown by $V_1$ and $V_2$. Equation (\ref{eqn_1}) relates the currents of the linear sub-circuit L and the excitation port to the other port voltages. Assuming the voltages are known at the beginning (this is what we call \textit{initial guess} later), the currents are found from the following matrix multiplication
\begin{equation}
\label{eqn_1}
\begin{bmatrix}\boldsymbol{I}_{L1} \\ \boldsymbol{I}_{L2} \\ \boldsymbol{I}_{L3}\end{bmatrix} = \begin{bmatrix} \boldsymbol{Y}_{11} & \boldsymbol{Y}_{12}  & \boldsymbol{Y}_{13} \\ \boldsymbol{Y}_{21} & \boldsymbol{Y}_{22} & \boldsymbol{Y}_{23} \\ \boldsymbol{Y}_{31} & \boldsymbol{Y}_{32} & \boldsymbol{Y}_{33} \end{bmatrix}\cdot\begin{bmatrix}\boldsymbol{V}_{1} \\ \boldsymbol{V}_{2}\\\boldsymbol{V}_{3} \\ \end{bmatrix} ,
\end{equation}
where $\boldsymbol{I}$ and $\boldsymbol{V}$ are vectors containing the Fourier components of the port currents and voltages, respectively. If the $k$ harmonics of the pump $\omega\pump$ are included in the Fourier spectrum of current and voltage, then each vector is $k+1$ long (by including the dc component or $\omega = 0$). Each vector is then written as follows:

\begin{equation}
\label{eqn_2}
\boldsymbol{I}_{Ln}=\begin{bmatrix}I_{Ln}(0) \\ I_{Ln}(\omega\pump)\\ \vdots \\ I_{Ln}(k\omega\pump)\end{bmatrix}, \boldsymbol{V}_{n}=\begin{bmatrix}V_{n}(0) \\ V_{n}(\omega\pump)\\ \vdots \\ V_{n}(k\omega\pump)\end{bmatrix}, n = 1, 2, ... .
\end{equation}

The components of the admittance matrix in equation (\ref{eqn_1}) form $k\times k$ diagonal matrices whose diagonal elements are values of $\boldsymbol{Y}$ (admittance) at pump harmonics. For example, $\boldsymbol Y_{31}$ is,
\begin{equation}
\label{eqn_3}
\boldsymbol{Y}_{31} = \begin{bmatrix} Y_{31}(0) & 0  & \ldots & 0 \\ 0 & Y_{31}(\omega\pump) & \ldots & 0  \\ \vdots & \vdots  & \ddots & 0 \\ 0 & 0 & 0 & Y_{31}(k\omega\pump) \end{bmatrix} .
\end{equation}
The current obtained from the nonlinear ports must satisfy Kirchhoff's current law (KCL). This means,
\begin{equation}
\label{eqn_4}
I_\mathrm{L} + I_\mathrm{NL} = 0. 
\end{equation}
The currents of the linear terminals, $\boldsymbol{I}_\mathrm{L1}$ and $\boldsymbol{I}_\mathrm{L2}$ are found from equation (\ref{eqn_5}), using the sub matrices of $\boldsymbol{Y}$. The part which is due to the contribution of the source (drive) voltage, $\boldsymbol{V}_3$, in the linear currents is shown by $\boldsymbol{I}\signal$. Compared to Fig. \ref{fig_1HB}, $\boldsymbol{I}\signal$ is the same as $\boldsymbol{I}_{N+1}$, where $\boldsymbol{N}=2$. The part of the admittance matrix which relates the port voltages to the currents is shown by $\boldsymbol{\Tilde{Y}}$.
\begin{equation}
\label{eqn_5}
\boldsymbol{I}_L = \begin{bmatrix} \boldsymbol{I}_{L1} \\ \boldsymbol{I}_{L2} \end{bmatrix} = \begin{bmatrix} \boldsymbol{Y}_{11} & \boldsymbol{Y}_{12}\\ \boldsymbol{Y}_{21} & \boldsymbol{Y}_{22} \end{bmatrix} \begin{bmatrix} \boldsymbol{V}_{1} \\ \boldsymbol{V}_{2} \end{bmatrix} + \boldsymbol{V}_3\begin{bmatrix} \boldsymbol{Y}_{13} \\ \boldsymbol{Y}_{23} \end{bmatrix} = \boldsymbol{\Tilde{Y}V} + \boldsymbol{I\signal}.
\end{equation}
The currents of the nonlinear terminals are found from:
\begin{enumerate}[label=(\alph*)]
    \item Inverse Fourier transform of the port voltages $\boldsymbol{V}_1$ and $\boldsymbol{V}_2$ to find $v_{1}(t)$ and $v_{2}(t)$,
    \item Finding the time-domain currents of the nonlinear sub-circuit terminals from the functions of the nonlinear elements (in this example two different or equal functions $g_1$ and $g_2$), and
    \item Finding the frequency-domain nonlinear currents using Fourier transform, $\mathcal F$.
\end{enumerate}

The summary of the above steps is written as
\begin{equation}
\label{eqn_6}
\boldsymbol{I}_{NL}=\mathcal F\begin{Bmatrix} i_{NL1}(t) = g_1(\mathcal F^{-1}\{\boldsymbol{V}_1,\boldsymbol{V}_2\}) \\ i_{NL2}(t) = g_2(\mathcal F^{-1}\{\boldsymbol{V}_1,\boldsymbol{V}_2\})\end{Bmatrix}.
\end{equation}

Then the following algebraic equation is formed by substituting equations (\ref{eqn_5}) and (\ref{eqn_6}) into equation (\ref{eqn_4}) and it is solved to find the new port voltage vectors, $\boldsymbol{V}$,

\begin{equation}
\label{eqn_7}
\boldsymbol{\Tilde{Y}V} + \boldsymbol{I}\signal + \boldsymbol{I}_{NL} = 0. 
\end{equation}

The left side of equation (\ref{eqn_7}) is a function of the voltage vectors, which is called current error $\boldsymbol{F(V)}$. The solutions for this equation are the points where the multidimensional surface $\boldsymbol{F(V)}$ crosses the coordinate axes. The number of crossing points is the number of common ports between the two sub-circuits (in this example $N=2$). While there are different algorithms to minimize the norm of the current error, the most common and powerful method to solve $\boldsymbol{F(V)}=0$ is to use the Newton-Raphson method \cite{Kundert2007,Gilmore91-1,Gilmore91_2}. This method is based on iterative solution of equation (\ref{eqn_7}) by starting from an initial guess for the solution of $\boldsymbol{F(V)} = 0$ (\eg, $\boldsymbol{V}_\text{old}$)
\begin{equation}
\label{eqn_8}
\boldsymbol{V}_\text{new}=\boldsymbol{V}_\text{new} + \left(\frac{\partial\boldsymbol{F}}{\partial\boldsymbol{V}}\right)^{-1}_{\boldsymbol{V}_\text{old}} \cdot \boldsymbol{V}_\text{old}.
\end{equation}

The new estimate of the solution, $\boldsymbol{V}_\text{new}$, is found by calculating the inverse of the Jacobian matrix in every iteration, which is,
\begin{equation}
\label{eqn_Jacobian}
\boldsymbol{J}=\partial\boldsymbol{F}/\partial\boldsymbol{V}.
\end{equation}

The size of this matrix is $2N(k+1)\times2N(k+1)$ where $N$ and $k$ are the number of ports and harmonics, respectively. The factor of two is for the imaginary and real parts of each port voltage at each harmonic.  
Details about convergence criteria, matrix solving algorithms, and methods for avoiding the local minima are discussed in \cite{Maas2003}.
Usually there are two different methods to calculate the inverse of the Jacobian matrix in equation (\ref{eqn_8}). The first method is a \textit{direct} method based on lower-upper (LU) factorization which is suitable for small circuits with a few nonlinear elements. In LU factorization, a square matrix, $\boldsymbol{A}$, is written as a product of a lower- and an upper triangular matrix, $\boldsymbol{A} = \boldsymbol{LU}$. The second method is the \textit{Krylov} sub-space method based on the generalized minimal residual (GMRES) method. The Krylov method is useful for circuits with a large number of harmonics or many nonlinear elements. Interested readers can find the details of the above algorithms in \cite{Golub2013}. \\
Choosing the maximum number of harmonics ($k$) is critical to obtain good convergence and correctly predict the circuit operations. A large value for $k$ makes the solution of equation (\ref{eqn_8}) very slow; on the other hand, a very small value for $k$ does not predict the real operation of the circuit.
An initial guess for equation (\ref{eqn_8}) is significant and must be judged based on the expected behavior of the circuit. For example, if the nonlinear circuit output has a diode which clips the signal, it is better to start with a clipped waveform as an initial condition instead of a full sinusoidal. 

Keysight's ADS uses the dc operating points of the circuit as a default initial guess. There is also an option  which is called transient-assisted harmonic balance (TAHB). With this option, the steady-state waveform resulting from a transient analysis is used as an initial guess for $\boldsymbol{V}$ in equation (\ref{eqn_8}). In this article, TAHB is used. Note that the time-step of the transient simulation should follow the Nyquist criterion which is in turn imposed by the maximum number of harmonics chosen for the pump (or large signal tone) in HB, $k$. 
Another option for an initial guess is to run an HB analysis with fewer harmonics and use the solution of this step for the second round of HB simulation with a larger number of harmonics. This is called HB-assisted HB or HBAHB. That means the solution of each HB analysis can be saved and reused as an initial guess for another round of HB simulations of the same or a similar circuit.

\section*{Appendix B}
\section*{Experimental Setup}
\begin{figure}[!t]
\centering
\includegraphics[width=3.5in]{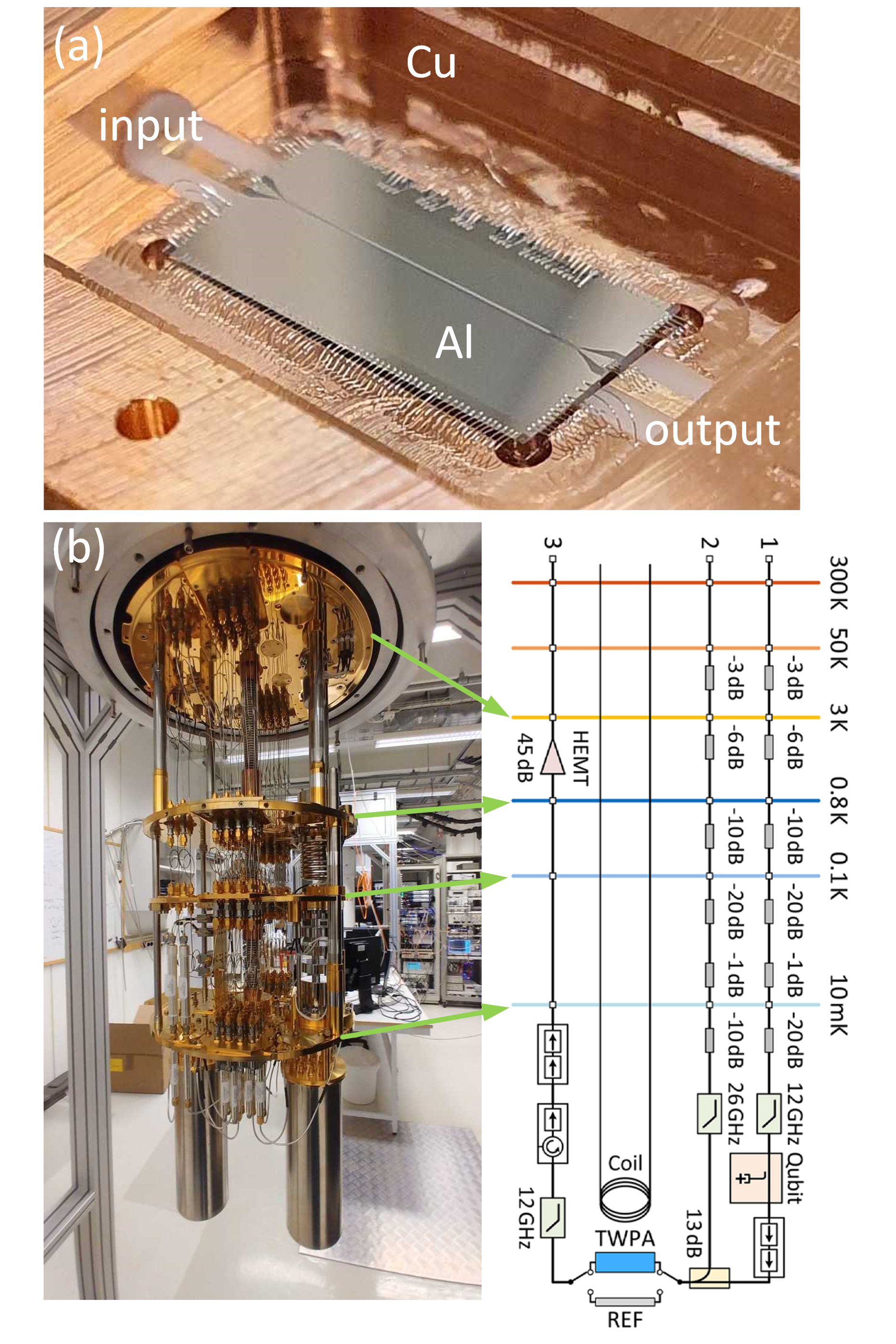}
\caption{(a) Bird's eye view of the TWPA chip installed in a copper package with two SMA connections at the input and output. (b) The cryogenic measurement setup and corresponding diagram with components used in each line.}
\label{ExpSetup}
\end{figure}

The TWPA chips are fabricated on high-resistivity (100) silicon wafers, and circuits are patterned on evaporated aluminum using nano-lithography and etching. The silicon chip is $7 \text{mm} \times 5 \text{mm}$, and its aluminum ground plane is wire bonded to the copper package. The TWPA chain resembles a CPW line [Fig. \ref{ExpSetup}(a)]. The packages are designed with special care to mitigate the package and chip modes. The TWPA is characterized at a cryogenic temperature of $10~\text{mK}$ within a dilution fridge [(Fig. \ref{ExpSetup}(b)] where the two-level energy splitting of qubits (typically $E=h\times4~\text{GHz}$) is immune from thermal noise and unwanted excitations using different attenuators and filters on lines 1 and 2. 
In the case of amplification of the qubit read-out signal, the signal is fed using line 1 in Fig. \ref{ExpSetup}(b), and the pump enters through line 2. They are combined before entering the TWPA using a coupler. The amplified signal is sent back to the room temperature stage via line 3 which has isolators and a HEMT amplifier at the 4K stage. In the case of characterizing the TWPA itself (\eg, measuring the gain spectrum), the qubit is bypassed, and signal and pump are combined at room temperature and then sent down to the TWPA via line 2. 
The $\unit[50]{\Omega}$ reference line is used to calibrate the attenuation of the line, as well as for a reference to obtain the gain of the TWPA. The hanging dc copper coil on the chip is used to apply external magnetic flux and tunes the bandwidth and the 3MW strength of the amplifier.

\bibliographystyle{IEEEtran}
\bibliography{Finalversion_HarmonicBalance_DShiri_etal_IEEEMWM_April2024}
\end{document}